%% file: apssamp.tex
\documentclass[%
 reprint,
nofootinbib,
 amsmath,amssymb,
 aps,
 prd,
]{revtex4-2}
\usepackage{array}
\usepackage{dcolumn}% Align table columns on decimal point
\usepackage{bm}% bold math
\usepackage{booktabs}
\usepackage{comment}
\usepackage{gensymb}
\usepackage{graphicx}% Include figure files
\usepackage{hyperref}% add hypertext capabilities
\usepackage{multirow}
\usepackage{orcidlink}
\usepackage{tabularray}
\usepackage{tabularx}
\usepackage{xcolor}

\newcommand{\MACSJ}{MACSJ0138}
\newcommand{\sigmalt}{\sigma_{\text{LT}}}
\newcommand{\sigmaap}{\sigma_{\text{ap}}}
\newcommand{\betasph}{\beta_{\text{sph}}}
\newcommand{\gammaext}{\gamma_{\text{ext}}}

\definecolor{jackgreen}{rgb}{0.0, 0.5, 0.0}

\definecolor{reviewerblue}{rgb}{0.0, 0.0, 0.5}

\begin{document}

\preprint{APS/123-QED}

%\title{Manuscript Title:\\with Forced Linebreak}% Force line breaks with \\
%\title{A Constraint on Dark Matter Self-Interaction\\from Combined Strong Lensing and Stellar Kinematics\\ in MACS J0138.0-2155}
\title{
A Constraint on Dark Matter Self-Interaction\\
from Combined Strong Lensing and Stellar Kinematics\\
in MACS J0138-2155}

\author{Jackson H. O'Donnell\orcidlink{0000-0003-4083-1530}}
 \email{jhodonnell@ucsc.edu}
\author{Tesla E. Jeltema\orcidlink{0000-0001-6089-0365}}
\author{M. Grant Roberts\orcidlink{0009-0005-7861-3290}}
\affiliation{%
 Department of Physics, University of California, Santa Cruz\\
 1156 High Street, Santa Cruz, CA 95064
}%
\affiliation{
Santa Cruz Institute for Particle Physics\\
1156 High Street, Santa Cruz, CA 95064
}

\author{James Nightingale\orcidlink{0000-0002-8987-7401}}
\affiliation{School of Mathematics, Statistics, and Physics, Newcastle University\\Newcastle, NE17RU}

\author{Abigail Flowers\orcidlink{0009-0001-7432-4361}}
\affiliation{
 Department of Physics, University of Z\"urich\\
 Z\"urich, Switzerland
}

\author{Dhruv Aldas\orcidlink{0009-0007-0103-1174}}
\affiliation{
 Department of Physics, University of California, Berkeley\\ CA 94720
}

\date{\today}% It is always \today, today,

\begin{abstract}
%We present a constraint on Self-Interacting Dark Matter (SIDM) in the galaxy cluster MACS J0138-2155, the host system of strongly lensed supernovae Requiem and Encore. This work robustly combines constraints from strong gravitational lensing with spatially resolved measurements of the central galaxy's dynamics, substantially improving on the methodology of previous SIDM constraints at cluster scales. The result is a self-consistent measurement of the density profile of MACS J0138-2155 across two orders of magnitude in radius. From the joint lensing and kinematics model, we report a $2\sigma$ (95\% confidence) upper limit on the SIDM cross section of $\sigma/m < 0.63$ cm$^2$/g, at a particle interaction velocity of $v_{\text{pair}} < $ {\color{red} TODO}. Furthermore, the methodology developed here advances the field of galaxy cluster strong lensing, and may inform future constraints on both SIDM interactions and time-delay measurements on the Hubble constant.
Self-Interacting Dark Matter (SIDM) represents a compelling alternative to collisionless dark matter, with diverse phenomenological signals from dwarf galaxy to galaxy cluster scales. We present new constraints on the SIDM cross section from the galaxy cluster MACS J0138-2155, host to the strongly lensed supernovae Requiem and Encore. Our analysis combines strong gravitational lensing with spatially resolved stellar kinematics of the central galaxy, employing several methodological advances over previous cluster-scale SIDM studies. The result is a self-consistent measurement of the density profile of MACS J0138-2155 across two orders of magnitude in radius. Our lensing and kinematics analyses individually yield highly consistent results, and from their combination we report a 95\% confidence upper limit on the SIDM cross section of $\sigma/m < 0.613$ cm$^2$/g, at an interaction velocity of $\langle v_\text{pair}\rangle < 2090$ km/s. This constraint, derived from the most detailed single-system analysis to date, is competitive with previous cluster-scale limits while demonstrating the power of combining complementary gravitational probes. The methodology developed here advances precision cluster lens modeling and will inform future studies of dark matter physics, as well as time-delay cosmography in this unique strong lensing system. Additionally, our results imply SN Requiem will reappear sooner than previously reported, with a $1\sigma$ CL between January 2027 and November 2028 at H$_0 = 67.7$ km s$^{-1}$ Mpc$^{-1}$.
%{\color{red} TODO Abstract}
%\begin{description}
%\end{description}
\end{abstract}

\maketitle

%\tableofcontents

\section{\label{sec:intro}Introduction}

%\reviewer{Overall, the authors could discuss more about the connections between the
%inferred SIDM physics and the models of the lens mass and kinematics. There is a
%brief discussion in appendix A but it is still unclear how the SIDM-predicted
%density profile is related to those observable.
%}

The nature of dark matter remains one of the most significant outstanding problems in physics.  An unambiguous, non-gravitational signature of dark matter has yet to be found in direct or indirect dark matter searches, nor have signs of beyond the Standard Model theories, like supersymmetry, containing dark matter particle candidates been seen. However, the growth and properties of cosmic structure place strong constraints on the nature of dark matter over a large range of scales.  The model of cold dark matter (CDM), in which the dark matter particles are collisionless, has been highly successful at modeling large-scale structure and its evolution, but faces potential tensions at smaller scales like dwarf galaxies.  This and a lack of detection of CDM particle candidates motivate thinking more broadly about the properties of dark matter.

Fortunately, non-CDM models for dark matter predict potentially measurable differences in the properties of dark matter structures, including their densities, shapes, and substructure abundances.  In particular, several models including self-interacting dark matter (SIDM), warm dark matter (WDM), and ultralight or ``fuzzy'' dark matter predict lower central densities in dark matter halos compared to CDM. All these alternatives to CDM reproduce its success at explaining large scale structures. WDM suppresses the linear matter power spectrum at small scales, accordingly limiting the growth of small scale dark matter halos. Axion-like or ``fuzzy'' dark matter can form soliton cores that in principle can affect the inner slope of the dark matter density profile, but on cluster scales the soliton core would be negligible compared to the size of the central galaxy and the main dark matter halo. On the other hand, SIDM can drive core-evolution and substructure on larger scales, which could impact lensing observables. Therefore, we limit our analysis to the case of SIDM. %\grantcomment{However, WDM will behave like CDM on large scales until free-streaming reduces the linear matter power spectrum and is constrained by Lyman-$\alpha$ observations.  } 

In this paper, we focus on using the combination of strong gravitational lensing (SL) and spatially resolved stellar kinematics to constrain the dark matter density to small radii in the galaxy cluster MACS J0138.0-2155, and we motivate these choices below.  In addition, the modeling presented here represents several improvements to the methodology and consistency of strong lens cluster modeling.

%{\color{red} DM is important and interesting, etc. Direct detection and indirect particle searches have not yet found DM. CDM successful at large scales, interesting small scale problems suggest extensions to CDM. Extensions to CDM affect the centers of DM halos, and possibly substructures. Strong lensing is a great way to look for these (many examples), especially in galaxy clusters. And kinematics is great too.}

\subsection{\label{sec:intro_sidm}Self-Interacting Dark Matter}

Self-Interacting Dark Matter has received intense interest in recent years, becoming one of the most well-studied alternatives to CDM. SIDM entails a simple phenomenological change to collisionless DM, in which DM particles are allowed to interact through a new force with a (possibly velocity-dependent) elastic scattering cross section. Such self-interactions are well-motivated theoretically, and arise naturally from many particle models of DM \cite{Tulin.Yu2018a,Adhikari:2022_SIDM}.
A common example is a massive DM particle coupled to a light mediator through a Yukawa interaction \cite{Tulin.Yu2018a,Feng2010:Yukawa,PhysRevLett.90.225002}.\footnote{We stress that our results do not assume a particular form of this particle interaction.}

SIDM can produce a range of different effects at different mass scales. In particular, scattering and associated heat transfer leads to the reduction of ``cuspy" DM density profiles into flattened ``cored" DM density profiles \cite{deBlok:core_cusp,Oman:2015xda}. %, but still having the freedom to produce diverse density profiles anywhere in between the two extremes 
On longer timescales, the gravothermal evolution caused by SIDM potentially leads to runaway gravothermal collapse. This evolution from core formation to core collapse naturally explains the diversity seen in dwarf galaxies; SIDM-induced core collapse could potentially even lead to the formation of supermassive black hole seeds in the early universe \cite{Outmezguine.Boddy.ea2022,Pollack_2015,2025JCAP...01..060G,2019JCAP...07..036C,Feng_2021}. %{JOD: \color{red} Potentially to add: Connection between subhalo core collapse and Meneghetti-type cluster lensing? The kind of subhalo core collapse that Gilman is interested in constraining?} \grantcomment{we could, though we'll have to spend a bit of time to motivate it. When you have merger's and non-relaxed environments, core-collapse gets a little wonky - it may or may not speed it up. Do you want me to look into it?}
At intermediate mass scales, the effects of SIDM on halo morphology are expected to be ``washed out" by baryonic effects \cite{Ragagnin.Meneghetti.ea2024,Mastromarino.Despali.ea2023}. Accordingly, most constraints on SIDM come from the extreme ends of the halo mass scale: dwarf galaxies, low-surface brightness (LSB) galaxies, and galaxy clusters \cite{Adhikari:2022_SIDM}. Observable signals could also exist at smaller scales, including detections of core collapsed low-mass haloes with gravitational imaging \cite{Li.Li.ea2025a,McKean.Spingola.ea2025,Vegetti.White.ea2026a}.

Currently, the most stringent absolute constraints on the SIDM cross section come from strong lensing of galaxy clusters, which suggest $\sigma/m \lesssim 0.1\,\text{cm}^2/\text{g}$ \cite{Sagunski.Gad-Nasr.ea2020,Andrade.Fuson.ea2022}. Constraints on dwarf and LSB mass scales from fitting rotation curves suggest a higher value, around $O(1-10)~\text{cm}^2/\text{g}$ \cite{Ren2019}. However, \citet{Roberts.Kaplinghat.ea2025} suggests that $\sigma/m$ could be as high as $20-40\,\text{cm}^2/\text{g}$ at this mass scale, when SIDM-induced core collapse is accounted for. These higher cross-sections alleviate tensions associated with the diversity of dwarf rotation curves by predicting that these low-mass galaxies are in different phases of their gravothermal evolution; most will be mildly in the core-collapsed phase, while others may only just be entering this phase, and some may be more deeply core collapsed. 

The discrepancy in the size of the cross-section implied by cluster and dwarf galaxy observations suggests that the SIDM cross-section must necessarily be a velocity-dependent interaction if we wish to explain dwarf rotation curves through this mechanism. It is clear that clusters can provide stringent constraints on the SIDM interaction strength at high mass scales. Therefore, to further probe the SIDM cross-section at these scales, in this paper we focus on strong lensing in galaxy clusters. 

%\grantcomment{Moved this discussion to the introduction}
%\reviewer{In addition, there are other alternative dark matter theories such as fuzzy
%dark matter, warm dark matter and axion dark matter etc. The authors could
%explain why this SIDM theory is more favorable than the others. 
%}

%\grantcomment{WDM kills structure too much, FDM/axions via soliton cores can maybe solve core-cusp - I'm not sure about diversity problem. Phenomenologically, axions require some more set up I think since they usually require being tied to the SM to solve the strong CP problem}

\subsection{Cluster SIDM constraints}

We aim to improve on previous SIDM constraints at cluster scales by constructing the most detailed model to date of an individual system. This work constrains the dark matter density across a wide range of radii by combining stellar kinematics of the central galaxy at small radii and observed SL features at intermediate radii. In this section we summarize previous related studies, and contrast the approach taken in this paper.

Early work along these lines aimed to differentiate between the luminous and dark matter distributions in clusters, independently using SL and stellar kinematics measured along a 1D slit \cite{Newman.Treu.ea2013a,Newman.Treu.ea2013b}, and found the DM profile alone to be shallower than expected in a CDM cosmology. Recently, \citet{Cerny.Jauzac.ea2025} took a similar approach, combining 2D IFU kinematics with SL to model six clusters, and likewise found a DM density profile shallower than a Navarro-Frenk-White (NFW) profile. However, these studies did not translate their measurements to SIDM constraints or any other alternate cosmological model. 

More recent studies explicitly constrained SIDM using similar approaches, albeit with less detailed models of larger samples. \citet{Sagunski.Gad-Nasr.ea2020} studied a sample of 8 galaxy groups and 7 galaxy clusters, constraining their density profile by combining stellar kinematics and strong gravitational lensing. Their results show a clear preference for non-zero cross-sections, with $\sigma/m = 0.19 \pm 0.09\,\text{cm}^2/\text{g}$ at cluster scales, and $\sigma/m = 0.5 \pm 0.2\,\text{cm}^2/\text{g}$ at group scales. For both kinematics and lensing, however, they limit their analysis to simplified approximations. For stellar kinematics, they use stellar velocity dispersions in bins along a 1D slit computed previously in \citet{Newman.Treu.ea2013b}, which yielded between 3 and 8 bins per cluster. In this work, we measure stellar velocity dispersions of a central cluster galaxy in 2D, using an IFU data cube to obtain 25 spatial bins, as described in Section~\ref{sec:stellar_kinematics_map}. Furthermore, their strong lens models are limited to reconstructing an average convergence in a certain aperture, derived from previous strong lens models of each system.

\citet{Andrade.Fuson.ea2022} presented an SIDM constraint from 8 lensing clusters, constructing a detailed SL mass model of each system, and relating those results to the SIDM cross-section $\sigma/m$. However, they use \textit{only} SL, without stellar kinematics to provide constraints on the innermost region. In addition, they choose a parametric cored-NFW density profile for the cluster DM halo, and relate their result to an SIDM cross-section after-the-fact. We improve on this latter choice by computing the SIDM halo density profile via the formalism described in Appendix~\ref{appendix:jeans}, and directly predicting strong lensing observables from this profile at every step.

We note that other probes exist to constrain SIDM in clusters, which do not explicitly reconstruct the DM density profile in the innermost region. These include offsets between the galaxies and dark matter in merging clusters \cite{Wittman.Stancioli.ea2023a}, offsets between the central galaxy and the center of mass \cite{Cross.Thoron.ea2024,Harvey.Robertson.ea2019a}, the radial acceleration relation of members in the cluster outskirts \cite{Tam.Umetsu.ea2023}, and potentially the ellipticity of dark matter haloes \cite{Robertson.Huff.ea2023,Gonzalez.Rodriguez-Medrano.ea2024}. Due to a combination of observational limitations and theoretical uncertainties, however, these constraints are not yet competitive with those from SL, such as \citet{Sagunski.Gad-Nasr.ea2020} and \citet{Andrade.Fuson.ea2022} described above.

\subsection{\label{sec:intro_system}Lensing Cluster MACS J0138.0-2155}

In this work, we focus on the strong lensing galaxy cluster MACS J0138.0-2155 (hereafter \MACSJ), whose primary source is a quiescent galaxy at z=1.95 known as MRG-0138 \citet{Newman.Belli.ea2018}. We leverage its deep, publicly-available data to constrain the inner density profile of its dark matter halo. This lensing cluster is of particular interest to cosmology: one of the first known strongly lensed supernovae, known as ``SN Requiem", was discovered in archival HST imaging of this system\cite{Rodney.Brammer.ea2021}. More recently, JWST observations of this system discovered a \textit{second} strongly lensed supernova, named ``SN Encore" \cite{Pierel.Newman.ea2024}, making this the first known lens system to host multiple strongly-lensed SNe. As a unique, highly interesting system, \MACSJ\,has been the subject of extensive astronomical observations. We describe the data used in this analysis in Section~\ref{sec:data}.

Furthermore, \MACSJ\, is a very ``clean" system, ideal for constraining its halo's inner density profile. It is known to be a relaxed, non-merging cluster, whose primary lensed source is a bright, quiescent galaxy at redshift 1.95. This source appears in at least 5 separate images, surrounding the cluster's center; furthermore, X-ray data show it to be a relaxed system, exhibiting close to spherical symmetry \citep{Flowers.ODonnell.ea2025}.

Here we briefly summarize previous studies on \MACSJ. Multiple spectroscopic studies have been done on the dynamical state of the cluster and its member galaxies \cite{Flowers.ODonnell.ea2025,Granata.Caminha.ea2025}, and the quiescent source hosting SNe Requiem and Encore \cite{Pierel.Newman.ea2024,Dhawan.Pierel.ea2024b,Newman.Belli.ea2018,Newman.Gu.ea2025}. The strong lensing observed in \MACSJ\,has been modeled previously by \citet{Newman.Belli.ea2018}, which reported spectroscopic redshifts of the lens and source, and \citet{Rodney.Brammer.ea2021}, which presented the discovery of SN Requiem. We confirm the redshifts presented in both papers using the IFU data described in Section~\ref{sec:data_muse}. More recent mass models leveraging deeper data on this system were published in \citet{Ertl.Suyu.ea2025} and \citet{Acebron.Bergamini.ea2025}. We compare the results presented in this work with the latter two papers in Section~\ref{sec:comparison_acebron_ertl}.

\section{\label{sec:data}Data}

The data used in this work includes space-telescope imaging and ground-based spectroscopy which are described below.

\subsection{\label{sec:data_imaging}High-resolution Imaging}

\MACSJ\, has received extensive high-resolution imaging from both HST and JWST. \MACSJ\, was first observed by HST in 2016 (Proposal ID 14496, PI A. Newman). These observations included three instrument/filter combinations: ACS/WFC band F555W, and WFC3/IR bands F105W and F160W. SN Requiem was later identified in these observations \cite{Rodney.Brammer.ea2021}. In 2019, \MACSJ\, was observed again by HST as part of the REQUIEM program (Proposal ID 15663, PI M. Akhshik). These observations utilized WFC3 exclusively, using the UVIS channel with filters F814W and F390W, and the IR channel with filters F110W, F125W, and F140W.
% See Rodney supplementary table 1 for detailed list of HST exposures

Following the discovery of SN Encore in November 2023, a JWST director's discretionary time program (JWST-GO-6549, PI J. Pierel) was approved to monitor SN Encore. This program observed \MACSJ\, in three epochs between December 5 2023, and January 8 2024. During each epoch, \MACSJ\, was observed by NIRCam using six filters: F115W, F150W, F200W, F277W, F356W, and F444W. We utilize the WFC3/IR imaging from Program 14496 to measure the observed positions of SN Requiem. To measure observed positions of other features in the lensed source, as well as SN Encore, we utilize JWST NIRCam imaging in bands F115W, F150W, and F200W from Program JWST-GO-6549.

\subsection{\label{sec:data_muse}MUSE Spectroscopy}

The Multi Unit Spectrographic Explorer (MUSE) is a second-generation integral field spectrograph on the VLT \cite{Bacon.Accardo.ea2010}. \MACSJ\, received three MUSE pointings of 970s each during Period 103, through Program 0103.A-0777(A). These observations used MUSE in wide-field mode, which has a field-of-view of approximately 1'x1', without adaptive optics. In this work, we utilize the reduced data cube of these observations obtained from the ESO data archive, processed by ESO pipeline version \texttt{muse/2.8} \cite{MUSEPipeline}.

\section{\label{sec:methods}Modeling Methods}

\begin{figure*}
    \centering
    \includegraphics[trim={0cm 4.5cm 0cm 1cm},clip,width=\linewidth]{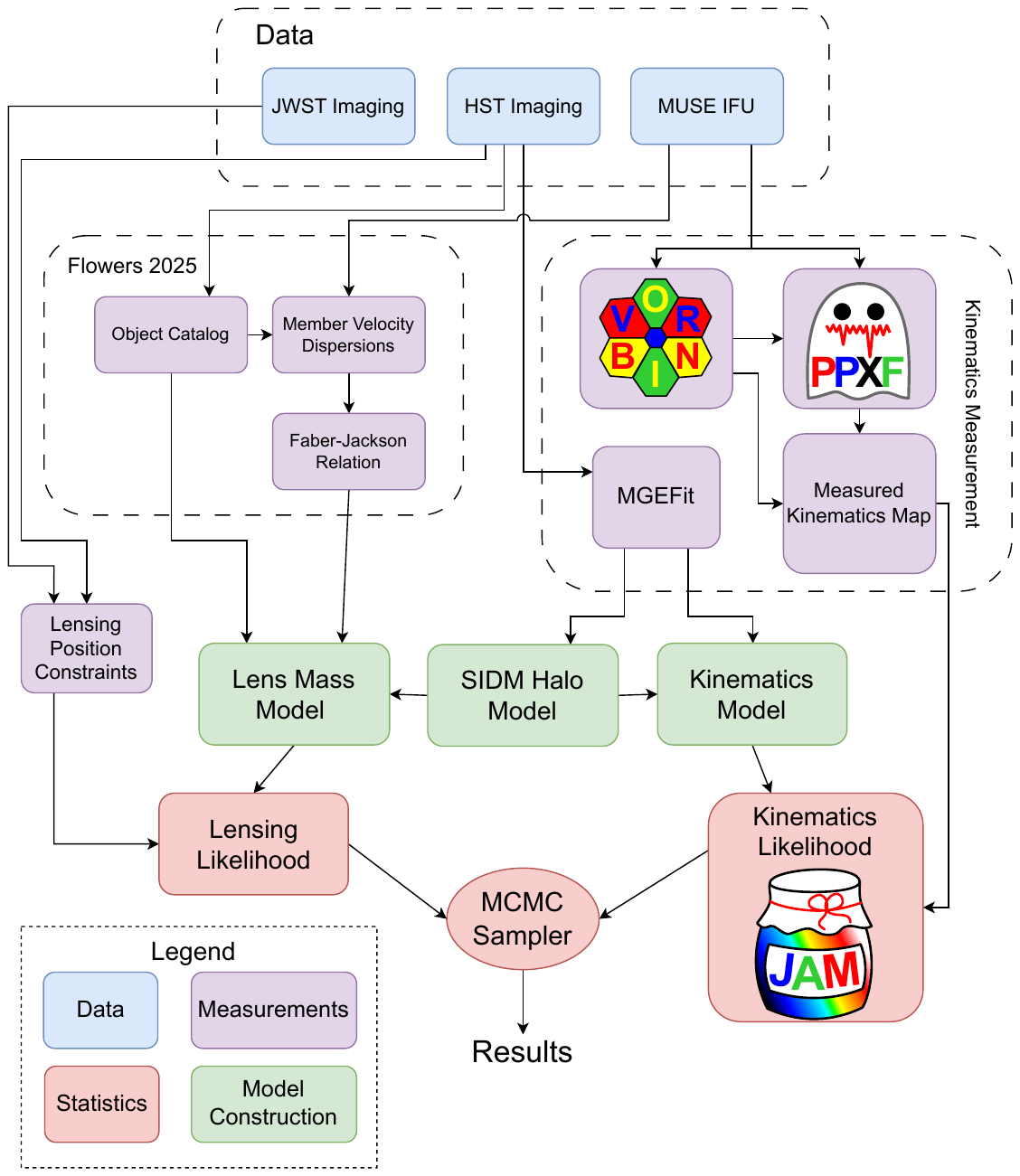}
    \caption{\textbf{Analysis Outline:} High-level overview of the analysis done in this work. Pieces of this analysis were presented in \citet{Flowers.ODonnell.ea2025}. Lens modeling is accomplished with \texttt{PyAutoLens} \cite{Nightingale2018,Nightingale2019}. Kinematics modeling in this work uses \texttt{vorbin} \cite{vorbin} to spatially segment the MUSE observations into regions, \texttt{ppxf} \cite{Cappellari.Emsellem2004,Cappellari2017} to model the spectrum of each region and extract stellar kinematics, \texttt{MGEFit} \cite{MGEFit} to produce a Gaussian expansion approximating the stellar luminosity of the BCG, and \texttt{JAM} \cite{Cappellari2008,Cappellari2012} to forward model the observed kinematics.
    %\reviewer{briefly explain those tools in terms of the physics and the Functionality}
    }
    \label{fig:analysis-flow-chart}
\end{figure*}

In this work, we self-consistently model and constrain the density profile of the dark matter halo of \MACSJ, using both stellar kinematics and strong lensing, combining complementary constraints of the DM density profile at $\sim 1-20$ kpc, and $\sim 100$ kpc, respectively. Central to this work is providing a coherent model for the 3D density of the overall DM halo, modified by the effects of SIDM. This shared mass model used by both the strong lensing and stellar kinematics probes is described in Section \ref{sec:methods_sidm_model}. Next, Section \ref{sec:methods_strong_lensing} describes how we construct a strong lensing mass model, and fit it to observed strong lensing features, including handling of the contribution of cluster member galaxies, and how strong lensing observables are computed for the SIDM-modified halo. Finally, the measurement and modeling of the central galaxy's kinematics are described in Section \ref{sec:methods_stellar_kinematics}. In addition, a high-level overview of the steps in this work is provided by the diagram in Figure~\ref{fig:analysis-flow-chart}.

\subsection{\label{sec:methods_sidm_model}Shared SIDM Mass Model}

We construct a self-consistent mass model of the dark and baryonic matter of \MACSJ\, across the regions probed by stellar kinematics and strong lensing. To represent the effect of self-interactions on the overall DM halo of \MACSJ, we utilize the isothermal Jeans method (see Appendix \ref{appendix:jeans} for details), as used by previous constraints on SIDM in clusters\cite{Andrade.Fuson.ea2022,Sagunski.Gad-Nasr.ea2020}
\footnote{We note that while this method only describes the `core-expansion' phase of SIDM, and fails to describe `core-collapse' behavior, current constraints already rule out SIDM-induced `core-collapse' at cluster scales. A simple estimate based on the relation in \citet{Outmezguine.Boddy.ea2022} predicts that a cluster like \MACSJ\, would experience `peak core' (i.e., the core-expansion/core-collapse phase transition) and begin collapsing after 216 Gyr of evolution at an SIDM cross section of 0.1 cm$^2/$g or 22 Gyr at a cross section of 1.0 cm$^2/$g.
%\grantcomment{this is probably not too important but by 'peak-core' which timescale do you mean precisely here again? the $t_{\rm coll}\sim~400 t_{c}$ one?}. \jackcomment{The $400 t_c$ is full core collapse. By `Peak core' I mean the minimum $\rho_0$.}\grantcomment{Ah I see, I would rephrase this to be something like 'experience the peak core size, i.e., the core-expansion/core-collapse phase transition'.}
}. % JOD: This estimate is for M200m = 10^(14.7), c200m = 8, at cluster redshift
In contrast to previous work such as \citet{Andrade.Fuson.ea2022}, which constrained a parametric density profile and only related this constraint to SIDM after-the-fact, we compute the SIDM-predicted density profile at every step and consistently model both stellar kinematics and strong lensing observables directly from this profile.
The SIDM-modified profile is predicted using the well-known isothermal Jeans formalism introduced in \cite{Vogelsberger.Zavala.ea2014}, as previous studies of SIDM in clusters have done \cite{Sagunski.Gad-Nasr.ea2020,Andrade.Fuson.ea2022}. This semi-analytic method consists of a simple modification of an NFW DM density profile, where a boundary is drawn at a spherical radius $r_1$, and all DM within $r_1$ is assumed to be isothermal. This boundary is conventionally defined as the radius at which DM particles have experienced, on average, one scattering over the lifetime of the halo:

\begin{equation}
    1 = \langle\frac{\sigma v}{m}\rangle \rho_\text{dm}(r_1) \, t_\text{halo}
\end{equation}

%\reviewer{The authors mentioned that the SIDM cross section $\sigma$/m is computed from the stellar mass-to-light ratio, M200m, c200m, and the radius $r_1$. Please clarify the actual steps to do this.}
%\reviewer{In Eq. (1), the halo age is mentioned but it is unclear if it is a constant or a function of time and SIDM physics. The discussion about the halo age could be added. }
Where $\langle\frac{\sigma v}{m}\rangle$ represents the velocity-weighted SIDM cross section per particle mass, and $t_{\text{halo}}$ is the age of the halo. We take the age of the halo to be the age of the universe at the redshift of the halo, which is 9.9 Gyr given our fiducial cosmology\footnote{This approach to $t_\text{halo}$ is the same as \citet{Andrade.Fuson.ea2022}. See Section 5.4 of \cite{Robertson.Massey.ea2021} for further discussion of how to treat $t_\text{halo}$.}. This simple formalism has been shown to effectively capture the `core formation' behavior of SIDM on cluster scales; see \cite{Robertson.Massey.ea2021} for a detailed study. Our implementation of the isothermal Jeans method is described in Appendix~\ref{appendix:jeans}, including the handling of priors on $r_1$ and $\sigma/m$, and the effect stellar density profile. Alongside this paper, we present our open-source code for computing the SIDM density profile with this method, available at \href{https://github.com/jhod0/sidm_halos}{\texttt{github.com/jhod0/sidm\_halos}}.

The baryon contribution in the innermost part of the halo is essential to both accurately computing the predicted SIDM profile, and predicting strong lensing and kinematics observables. In this work we consider the impact of the stellar mass of the central galaxy of the cluster (typically the Brightest Cluster Galaxy, or BCG). The light profile of the BCG is approximated as a Multi-Gaussian Expansion (MGE), as described in Section \ref{sec:stellar_kinematics_mge}, and the baryonic mass is derived with a stellar mass-to-light ratio. We use \texttt{kcorrect}\cite{kcorrect} version 5.1.3 to compute the V-band absolute magnitude of the BCG using Legacy Survey photometry in the g, r, i, and z bands \cite{LegacySurvey}; in addition, \texttt{kcorrect} provides an estimate of the stellar mass-to-light ratio $\Upsilon_*$. This yields a V-band absolute luminosity of $4.94 \times 10^{11} \text{L}_\odot$ with a corresponding $\Upsilon_{*,V}$ of $2.05$. We note this is similar to the result in \citet{Newman.Treu.ea2013a}, which similarly studied BCGs of lensing clusters and found $\Upsilon_{*,V}$ between 1.8 and 2.32 across seven systems. It is well known that there is considerable uncertainty on the true mass-to-light ratio, which is highly dependent on the assumed initial mass function (IMF) \cite{Sagunski.Gad-Nasr.ea2020,Newman.Treu.ea2013b}. As such, and following \citet{Sagunski.Gad-Nasr.ea2020}, we use a wide prior on $\log\Upsilon_*$ with a scatter of 0.3 dex.

Additionally, we allow for a prolate overall DM halo in three dimensions, rather than simply ellipticity in the plane of the sky.  This treatment of the ellipticity introduces three parameters: the prolate axis ratio $q_{3d}$, the angle between the halo's axis of symmetry and the line of sight $\cos i$, and a position angle $\phi$ of the apparent ellipticity in the plane of the sky. We introduce ellipticity to the overall halo while ensuring that the mass and concentration of the halo are preserved, and analytically project this prolate halo into an elliptical surface density in the plane of the sky. A detailed explanation can be found in Appendix~\ref{appendix:triaxiality}.

\begin{figure*}
    \centering
    %\includegraphics[width=\linewidth]{}
    %\framebox(\linewidth,0.5\linewidth){Placeholder}
    \includegraphics[trim={1cm 3.5cm 2cm 4.5cm},clip,width=\linewidth]{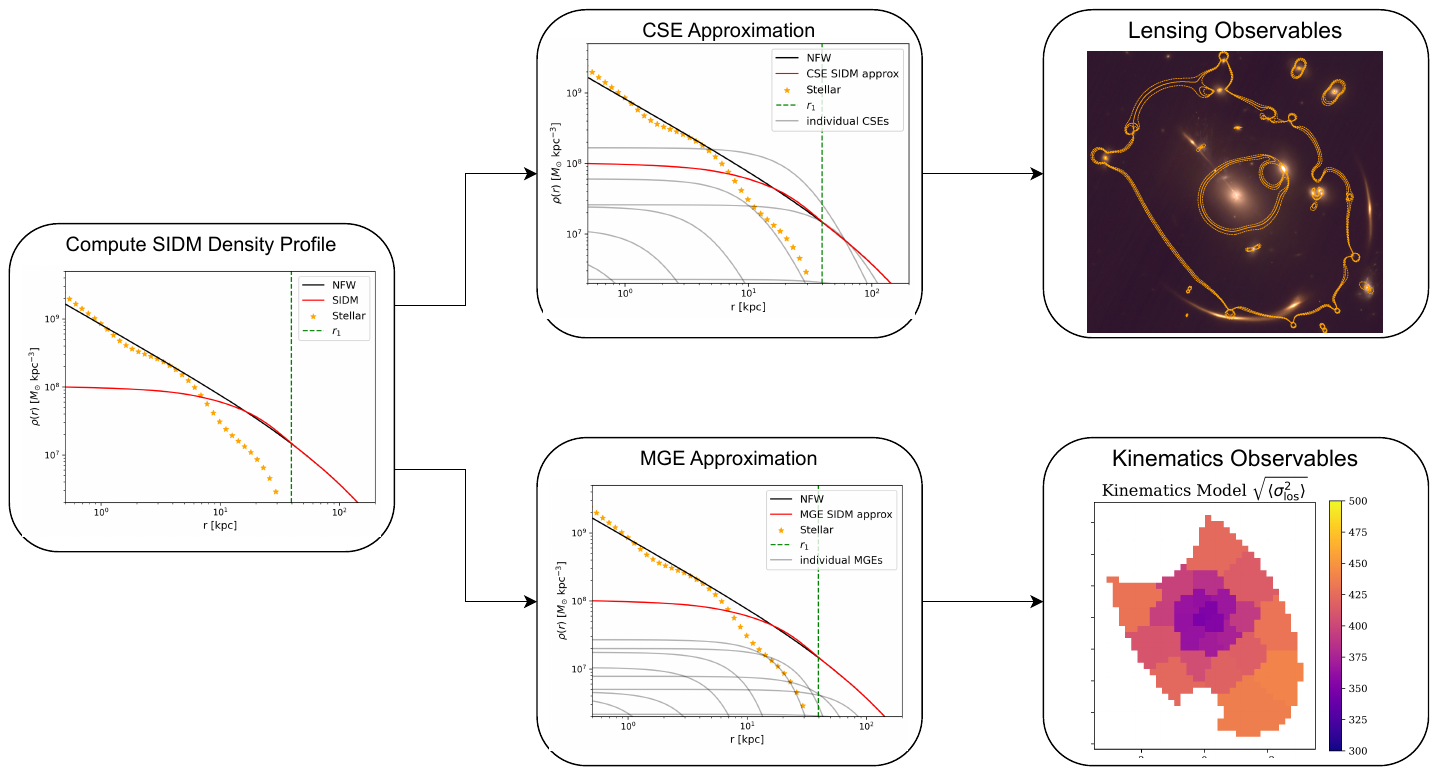}
    \caption{\textbf{SIDM Analysis:} Illustration of how the SIDM-modified dark density profile is translated into lensing and kinematics observables. In all figures, an unmodified NFW profile is shown for reference in black, and the SIDM-predicted dark density profile is shown in red. \textbf{Left:} First, the isothermal Jeans equation is solved to compute the predicted SIDM-modified density profile, as described in Section~\ref{sec:methods_sidm_model} and Appendix~\ref{appendix:jeans}. \textbf{Middle:} The SIDM-modified density profile is approximated by a sum in a particular basis suitable for each observable. For lensing, we use CSEs, and for kinematics we use an MGE. In both cases, the approximation is accurate to within 1.5\% across the radii shown. \textbf{Right:} The model is fit to data. See Sections~\ref{sec:methods_strong_lensing} and ~\ref{sec:methods_stellar_kinematics} for details.}
    \label{fig:profile-decomp}
\end{figure*}

In order to compute lensing and kinematic observables from this prolate, SIDM-modified density profile, the density profile is first approximated by a basis sum appropriate for each observable. The lensing observables are computed with Cored Steep Ellipsoids (CSEs), as described in Sec.~\ref{sec:methods_strong_lensing}, and the kinematic observables are computed with a Multi-Gaussian Expansion (MGE), as described in Sec.~\ref{sec:stellar_kinematics_jam}. A visual summary of this process is presented in Figure~\ref{fig:profile-decomp}.

%{\color{red} Shared model now also includes: $q_{3d}$, $\cos(i)$, $\phi$. TODO explain our new ellipticity formalism, and how all these come together.}

\subsubsection{\label{sec:methods_sidm_parameters}Shared Model Parameters}

There are thus seven parameters common to both the strong lensing and kinematics models. Together, they describe the overall dark matter halo and inner baryon density profile. They are:

\begin{itemize}
    \item \textbf{$\log_{10}M_{200m}$:} The overall mass of the cluster's DM halo, defined as the mass within radius $r_{200m}$ such that the mean density within $r_{200m}$ is 200 times the mean matter density of the universe;
    \item \textbf{$c_{200m}$:} The corresponding concentration of the cluster's DM halo;
    \item \textbf{$\log_{10}r_1$:} The radius within which DM particle self-interactions modify the DM density profile; \footnote{Though we sample $r_1$ rather than $\sigma/m$, likelihoods are weighted to enforce a flat prior on $\ln(\sigma/m)$. See Appendix \ref{appendix:jeans}.}
    \item \textbf{$\Upsilon_*$:} The stellar mass-to-light ratio of the central galaxy;
    \item \textbf{$q_{3d}$:} The 3D axis ratio of the cluster halo;
    \item \textbf{$\cos(i)$:} The \textit{cosine} of the inclination angle relative to LOS;
    \item \textbf{$\phi$:} The position angle of the projected, 2D ellipticity on the sky.
\end{itemize}

\subsection{\label{sec:methods_strong_lensing}Strong Lens Modeling}

Lens modeling in this work is done with the flexible open-source package \texttt{PyAutoLens} \cite{pyautolens,Nightingale2018,Nightingale2019}. In addition to the built-in functionality of \texttt{PyAutoLens}, we implement several extensions to describe our lens model and constraints. The mass model used in this work, and the functionality implemented to enable it, are described in Section~\ref{sec:lensing_mass}; the strong lensing constraints, and the likelihood implemented in this work, is described in Section~\ref{sec:lensing_constraints}.

\subsubsection{\label{sec:lensing_mass}Mass Model}

Our strong lensing mass model takes a similar approach to the widely-used \texttt{lenstool} software \cite{Jullo.Kneib.ea2007,Kneib.Bonnet.ea2011}. It contains two primary components: one dark matter halo for the cluster as a whole, and a set of dPIE\cite{Eliasdottir.Limousin.ea2007} profiles representing all member galaxies, save for the central galaxy.

To represent the former, we approximate our SIDM-modified DM density profile via a sum of Cored Steep Ellipsoids (CSEs), as proposed by \citet{Oguri2021}. This CSE decomposition is computed by the \texttt{scipy} method \texttt{least\_squares}, using the NFW approximation in Table 1 of \citet{Oguri2021} as an initial value. The position of the overall halo is allowed to vary slightly, with a Gaussian prior centered on the known BCG center.

In order to model the member galaxies, we implement a dPIE density profile in \texttt{PyAutoLens}. The mass of these member galaxies are all scaled by their photometry using the Faber-Jackson relation, described below; we additionally leverage a direct measurement of the Faber-Jackson relation in \MACSJ, as advocated by \citet{Bergamini.Rosati.ea2019}. This measurement was previously reported in \citet{Flowers.ODonnell.ea2025}, yielding a slope of $\alpha \sim 0.26 \pm 0.06$ at a reference velocity dispersion of $\sim 220$ km/s. We note that while this measurement was done using the older, shallower MUSE data, an independent team reported a similar measurement of the Faber-Jackson relation in \MACSJ\, using the newer, deeper MUSE observations. These results, published in \citet{Granata.Caminha.ea2025}, used different methodology on different data, yet yielded the highly consistent measurement of $\alpha \sim 0.25 \pm 0.05$. Both results suggest a slightly shallower slope than those reported for three clusters in \citet{Bergamini.Rosati.ea2019}.

The dPIE profiles of the member galaxies are parameterized by their velocity dispersion $\sigma_i$, core radius $r_{a,i}$, and scale (or `cut') radius $r_{s,i}$. Except where otherwise noted, all member galaxies receive a fixed core radius of 
% Red-sequence members: core 0.15kpc
% Jellyfish: 0.1kpc
0.15 kpc. Their velocity dispersions and cut radii are scaled in the conventional way, as in \cite{Jullo.Kneib.ea2007} and \cite{Bergamini.Rosati.ea2019}:

%{\color{red} TODO JACK CHECK: modulo LT v. disp. versus "true" v. disp. See Bergamini section 4.}
\begin{align}
    \sigma_{\text{LT},i} &= \sigma_{\text{LT},\text{ref}}\Big(\frac{L_i}{L_0}\Big)^\alpha \\
    r_{s,\text{LT},i} &= r_{s,\text{LT},\text{ref}} \Big(\frac{L_i}{L_0}\Big)^\beta
\end{align}

Here, LT denotes the fiducial parameterization used in \texttt{lenstool}, rather than observed quantities. If the total mass-to-light ratio $\Upsilon^{\text{tot}}_i$ for these galaxies scales as $L_i^\gamma$, then $\beta = \gamma - 2\alpha + 1$. Following both \citet{Bergamini.Rosati.ea2019} and \citet{Acebron.Bergamini.ea2025}, we fix $\gamma = 0.2$. The fiducial dPIE velocity dispersion, $\sigmalt$, is known to differ from observed aperture-averaged velocity dispersions $\sigmaap$; in the limit $r_a \rightarrow 0, r_s \rightarrow \infty$, the dPIE becomes an SIS, and $c_p = \sigmaap/\sigmalt = \sqrt{3/2}$ \cite{Eliasdottir.Limousin.ea2007}. In practice, $c_p$ is generally closer to unity. \citet{Bergamini.Rosati.ea2019} provides a practical treatment of $c_p$ in three lensing clusters with measured Faber-Jackson relations, they find that for all three clusters, given physical values of $r_a$ and $r_s$, $c_p \approx 1.12$. For our priors on the substructure scaling relations in \MACSJ, we therefore adjust the measured Faber-Jackson relation from \cite{Flowers.ODonnell.ea2025} by $c_p = 1.12$.

\begin{figure}
    \centering
    \includegraphics[width=\linewidth]{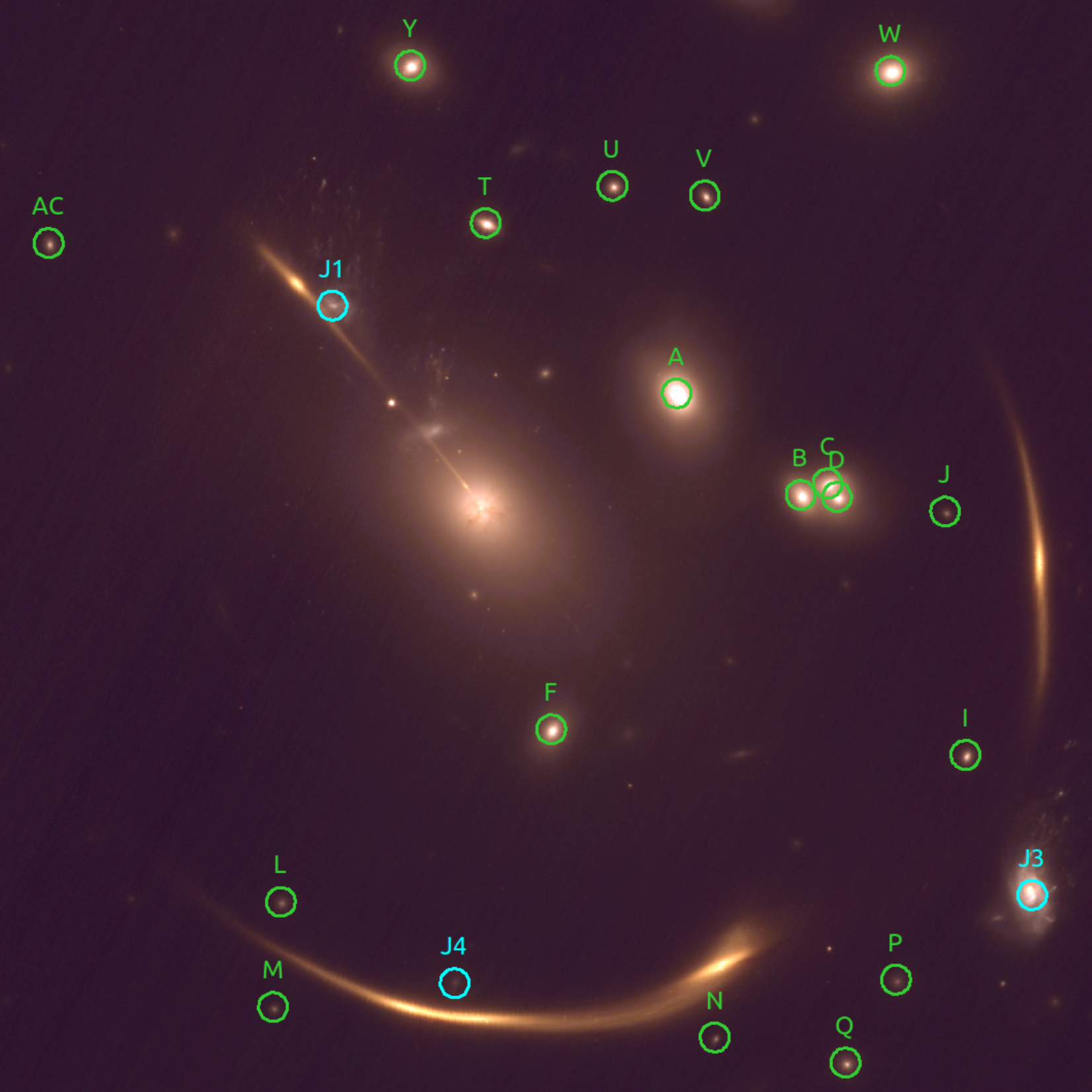}
    \caption{JWST NIRCam imaging of \MACSJ. All galaxies labeled here included as components of the strong lensing mass model. Red-sequence galaxies are shown in green, emission-line galaxies are in cyan. All circles have radius 0.5''. Several cluster members outside this FOV are also included in the mass model. In this RGB image, the blue, green, red channels show the F115W, F150W, and F200W NIRCam filters, respectively.}
    \label{fig:members}
\end{figure}

Red-sequence cluster members included in the mass model are chosen from the catalog introduced in \citet{Flowers.ODonnell.ea2025}, all of which are spectroscopically confirmed to lie at or near the cluster's redshift.\footnote{There exists a small subset of red-sequence galaxies at higher redshift, with a velocity offset of $\sim8,000$ km/s relative to \MACSJ\, \cite{Flowers.ODonnell.ea2025,Granata.Caminha.ea2025}. } In addition, several emission-line galaxies near the cluster redshift lie close to the observed lensing features, and their mass contributions are necessary to construct a precision mass model. We include three of these in our mass model, as shown in Figure \ref{fig:members}; note that we label these following the convention in \citet{Gibson2025}, which differs from that in \citet{Rodney.Brammer.ea2021} and \citet{Acebron.Bergamini.ea2025}. The fourth galaxy studied in \cite{Gibson2025}, J2, is not included in our mass model: its projected position is so close to the cluster center that it's contribution to the observed lensing is dwarfed by the BCG and cluster DM halo.

Our lensing model includes one further component, a constant `external shear' $\gammaext$ across the \MACSJ\, field. This shear is parameterized by the two conventional components $(\gamma_1, \gamma_2)$, where $\gamma_1$ shears in the vertical and horizontal directions, and $\gamma_2$ shears along the diagonals. \citet{Acebron.Bergamini.ea2025} likewise includes a $\gammaext$ component to improve their mass model, however \citet{Ertl.Suyu.ea2025} models \MACSJ\, using two halo components without $\gammaext$. A discussion of our $\gammaext$ result and its interpretation can be found in Section~\ref{sec:results_external_shear}.

%{\color{red} Lensing adds 10 (!) params: $\gamma_{1,2}$, $x/y_{halo}$, $\sigmalt^{\text{piv}}$, $\alpha$, $r_s^{\text{J4}}$, $\sigmalt^{J1,J3,J4}$.}

\subsubsection{\label{sec:lensing_constraints}Lensing Constraints}

Our lensing mass model is constrained by a variation of the conjugate point method, computed in the source plane. For each observed source feature $F$, observed at multiple positions $\theta^F_{i}$, the image-plane uncertainties are ellipsoids described by the covariance matrix $\Sigma_i^F$. For a given mass model, the corresponding source-plane positions and covariance are:

\begin{align}
    \beta_i^F &= \theta_i^F - \alpha(\theta_i^F) \\
    B_i^F &= A(\theta_i^F) \Sigma^F_i A^T(\theta_i^F)
\end{align}

Where $A(\theta)$ is the lensing Jacobian at image plane position $\theta$. Given these predicted source plane positions, we analytically marginalize over the unknown intrinsic source plane position for each feature. In practice, to avoid overfitting to critical curves of the lens model, we average $B_i^F$ across all image-plane positions within the uncertainty ellipsoid described by $(\theta_i^F, \Sigma_i^F)$. For each set of observed features $F$, this yields a best-fit source plane position $\hat\beta^F$, with corresponding uncertainty ellipsoid $\hat B^F$, and a likelihood $\mathcal{L}^F$:

\begin{align}
    \hat\beta^F =& \hat B ^F\big(\sum_i(B^F_i)^{-1}\beta_i^F\big)\\
    (\hat B^F)^{-1} =& \sum_i (B^F_i)^{-1}\\
    \begin{split}
    \log\mathcal{L}^F =& -\log \sqrt{\frac{|\hat B^F|}{2\pi}} \\
    &- \frac{1}{2}\sum_i (\beta_i^F - \hat\beta^F)^T(\hat B^F)^{-1}(\beta_i^F - \hat\beta^F)
    \end{split}
%\end{equation}
\end{align}

This likelihood is implemented as an extension of \texttt{PyAutoLens}. This method was later found to be equivalent to that introduced in Section 5.1 of \citet{gravity.jl}.

\begin{table*}[]
    \centering
    \input{position_constraints_table}
    \caption{\textbf{Position constraints:} The constraints used for the strong lensing mass model. Each group of observed features is given a separate label (e.g. `SN-E', `A'), and each separately observed position is given a number (`SN-E.1', `A.2'). Position angles are given east-of-north.}
    \label{tab:position_constraints}
\end{table*}

\begin{figure*}
\includegraphics[width=1.0\linewidth]{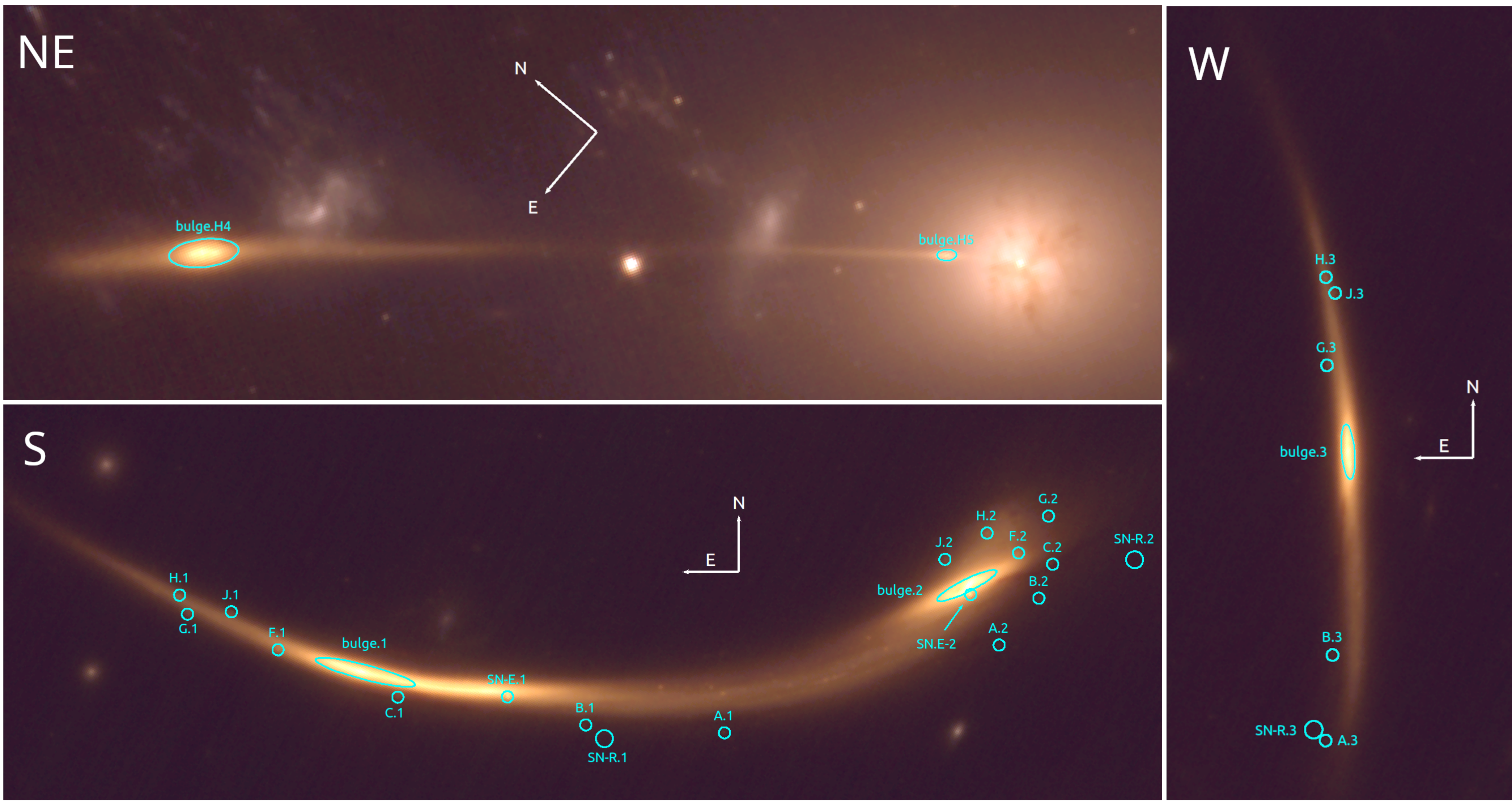}
    \caption{\textbf{Lensing position constraints:} The positions of lensed features used to constrain the strong lens model, using the likelihood described in Section \ref{sec:lensing_constraints}. The RGB color bands and scaling are the same as Figure \ref{fig:members}. Three fields are shown here, showing lensed features of the quiescent source in the South, West, and North-East. In each field, a compass in displayed showing North and East; both segments of the compass are 1" long for scale.}
    \label{fig:position_constraints}
\end{figure*}

Our constraints include features in all five arcs of the primary quiescent source. Although recent work has reported additional sources from faint emission lines in deep MUSE data\cite{Granata.Caminha.ea2025}, those sources are not used in this work. All position constraints are measured from JWST NIRCam imaging, save for three observed images of SN Requiem, which was only seen at an earlier epoch. The positions of the three SN Requiem images are taken from \citet{Rodney.Brammer.ea2021}, and assigned a higher position uncertainty due to the lower resolution of HST imaging. The position constraints used here are shown in Figure \ref{fig:position_constraints}, and listed in Table \ref{tab:position_constraints}.

\subsection{\label{sec:methods_stellar_kinematics}Stellar Kinematics of the BCG}

Stellar kinematics of SL deflectors provides a powerful, independent probe of the lensing mass distribution, especially if it can be resolved spatially. Early work along these lines used slit spectra of central cluster galaxies to measure variations in the stellar velocity dispersions in one dimension \cite{Sand.Treu.ea2004,Newman.Treu.ea2013a,Newman.Treu.ea2013b}. Much recent work has leveraged IFU spectra to study the dynamics of lensed systems, including lensed quasar systems \cite{Shajib.Mozumdar.ea2023,Knabel.Treu.ea2024a,Knabel.Mozumdar.ea2025}, time delay constraints on the Hubble constant \cite{Birrer.Buckley-Geer.ea2025,Shajib.Treu.ea2025}, galaxy-scale multi-source-plane lenses \cite{Turner.Smith.ea2024a}, and even a group-scale strong lens \cite{Wang.Canameras.ea2022a,Wang.Canameras.ea2024a}. In the majority of these studies, the primary contribution of the stellar dynamics is to refine the strong lensing model, yielding a more accurate result and potentially breaking the mass-sheet degeneracy. In our case, however, the inner density profile constrained by central galaxy's dynamics is of direct scientific interest: SIDM affects the halo's dark matter density in precisely this region.

Here we describe the procedure for measuring spatially resolved stellar kinematics, and how it is used to constrain our mass model. Sections \ref{sec:stellar_kinematics_map} and \ref{sec:stellar_kinematics_cov} describe how the kinematics map was measured and validated; Section \ref{sec:stellar_kinematics_mge} describes the light profile of the central galaxy, essential to modeling the kinematics; and Section~\ref{sec:stellar_kinematics_jam} describes how our mass model is constrained by this data.

\begin{figure*}
    \centering
    \includegraphics[width=0.57\linewidth]{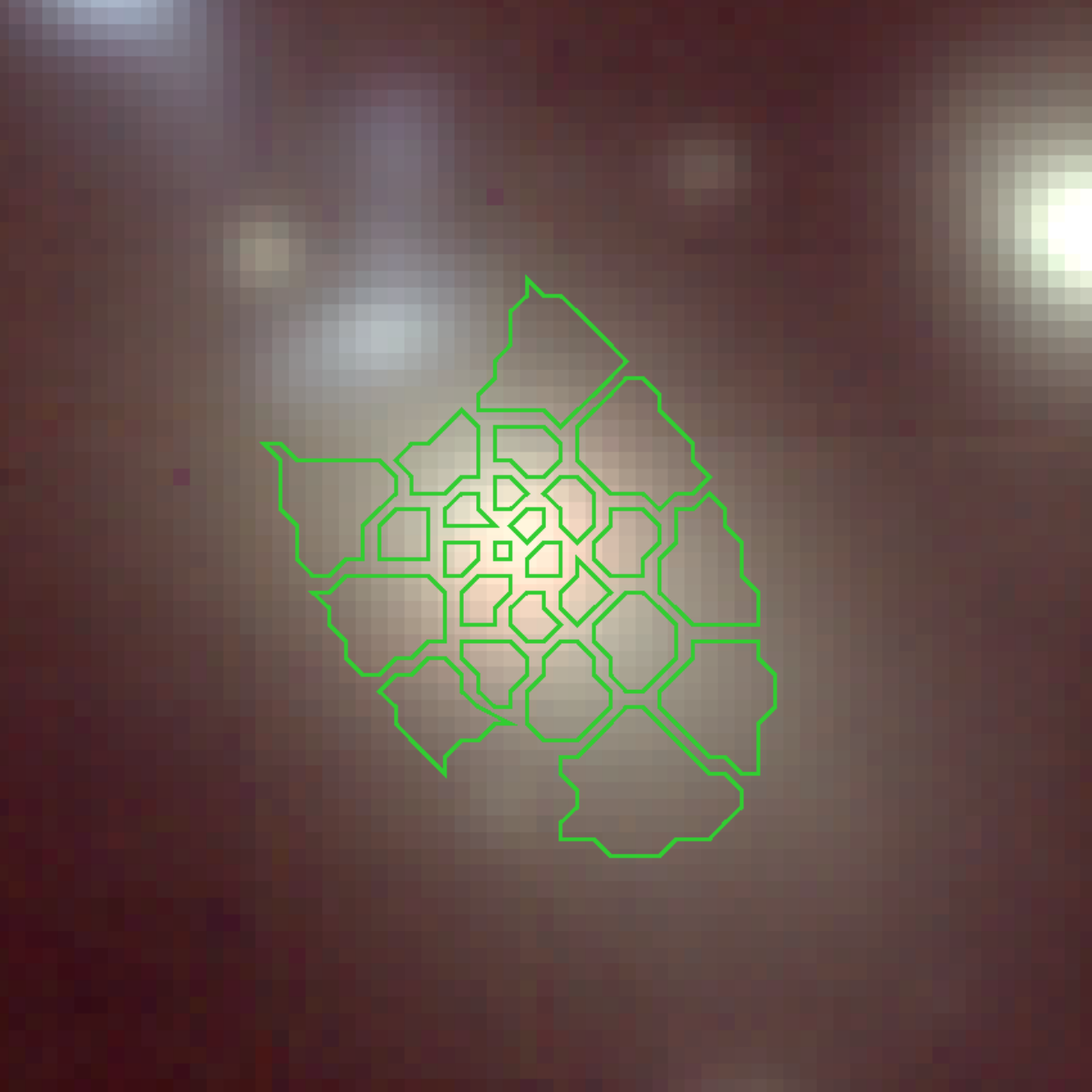}
    \includegraphics[width=0.34\linewidth]{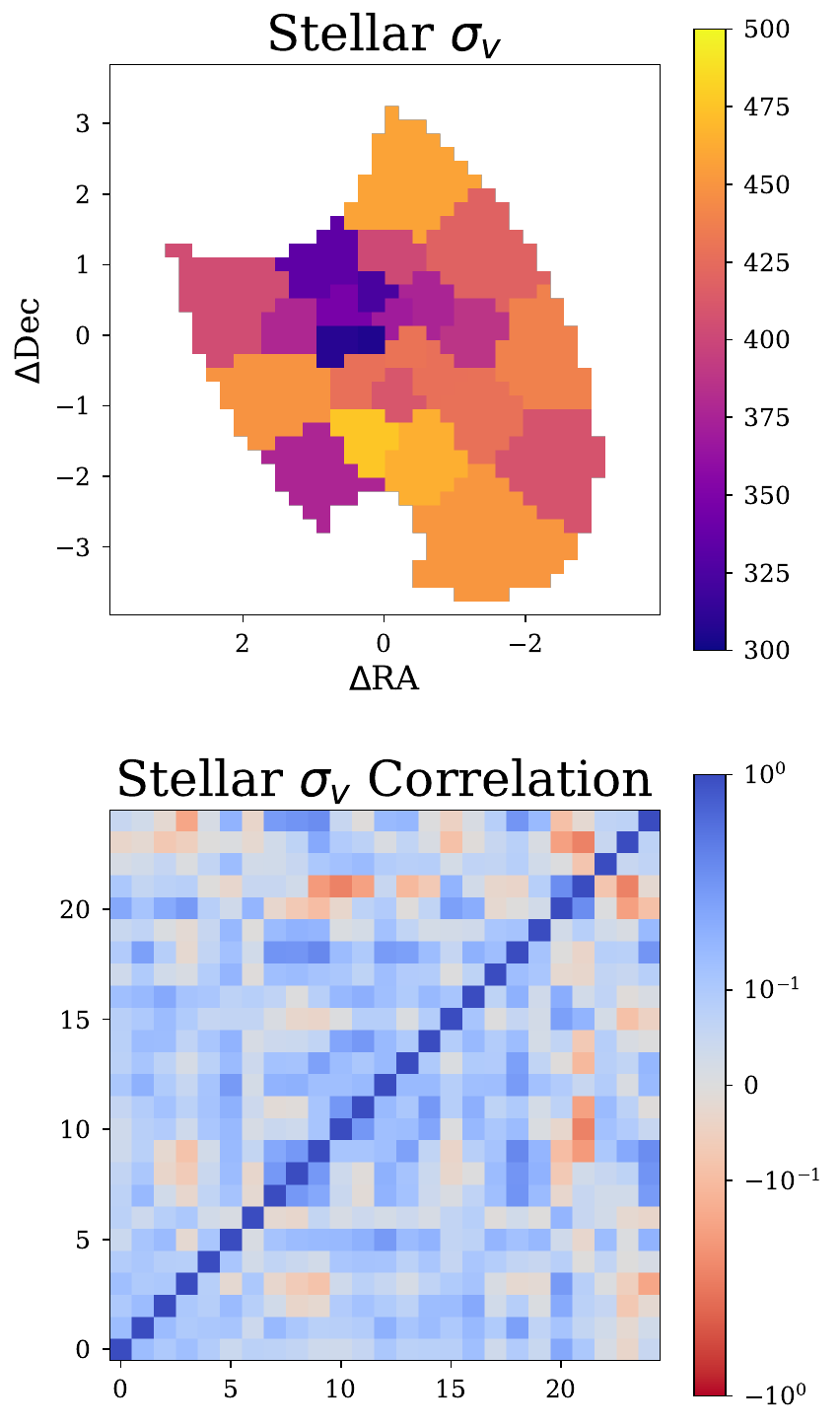}
    \caption{The central galaxy of \MACSJ, and the LOS stellar velocity dispersions measured in this work. \textbf{Left:} MUSE color imaging of the central galaxy, with the Voronoi bins used in this work overlaid in green. \textbf{Upper Right:} The LOS stellar velocity dispersion in each bin, in km/s. \textbf{Lower Right:} The correlation matrix of the LOS velocity dispersion in all 25 spatial bins, displayed in a \texttt{symlog} scaling with a linear threshold of 0.2.}
    \label{fig:kinematics-measurement}
\end{figure*}

\subsubsection{\label{sec:stellar_kinematics_map}Measuring the Kinematics Map}

To robustly measure the stellar kinematics in numerous spatial bins across the central galaxy, we take a similar approach to \citet{Shajib.Mozumdar.ea2023}. We consider all IFU `spaxels' along the central galaxy, binning them spatially using \texttt{vorbin}\cite{vorbin} such that each bin attains a target spectral signal-to-noise. We use all pixels within an ellipse tracing the BCG, with a major axis of 8.5" and a minor axis of 5". Two circular regions are excluded due to contamination by other galaxies along the line of sight. These spectra are spatially binned to a target signal-to-noise of 23/\AA\,between the rest-frame wavelengths of 5035 and 5160\AA, just below the Mg b absorption feature. This yields 25 spatial bins, as shown in Figure~\ref{fig:kinematics-measurement}.

Spectral fitting is done with the widely used software package \texttt{ppxf}\cite{Cappellari.Emsellem2004}. This tool requires a set of input templates, and an optimal combination of templates is found which closely represents the user's spectrum. In this work, we use the X-Shooter Library (XSL) Data Release 3\cite{Verro.Trager.ea2022}, a set of 830 spectra of 683 diverse stars, as our template library. %We apply several cuts to this library to exclude low-quality spectra.
All template spectra are normalized by their mean flux between 4000\AA\, and 4200\AA. To exclude low-quality spectra, the mean SNR per pixel in this region is computed, and template spectra with SNR less than one are excluded. This is similar in spirit to the ``cleaning'' advocated by \citet{Knabel.Mozumdar.ea2025}.

We compute maps of the stellar kinematics in each Voronoi bin as follows. First, the spectrum of the entire BCG is extracted (i.e., all IFU `spaxels' in all Voronoi bins). Then \texttt{ppxf} is run once on this total spectrum, and provided as templates the XSL stellar spectra described above. An example of this result is shown in Fig~\ref{fig:ppxf_fits}. This fit determines a linear combination of XSL template spectra which most closely matches our observed spectrum; we record this combination as the \textit{optimal template}. Next, for each Voronoi region, \texttt{ppxf} is applied to the region's spectrum, this time using the single optimal template for the stellar component rather than providing a library of templates. For each spectral fit, we allow \texttt{ppxf} to fit velocity moments to the 4th order, meaning we allow for nonzero values of the third- and fourth-order moments $h_3$ and $h_4$\footnote{See \citet{Cappellari.Emsellem2004} for an explanation of these moments.}.

\begin{figure*}
    \centering
    \includegraphics[width=1.0\linewidth]{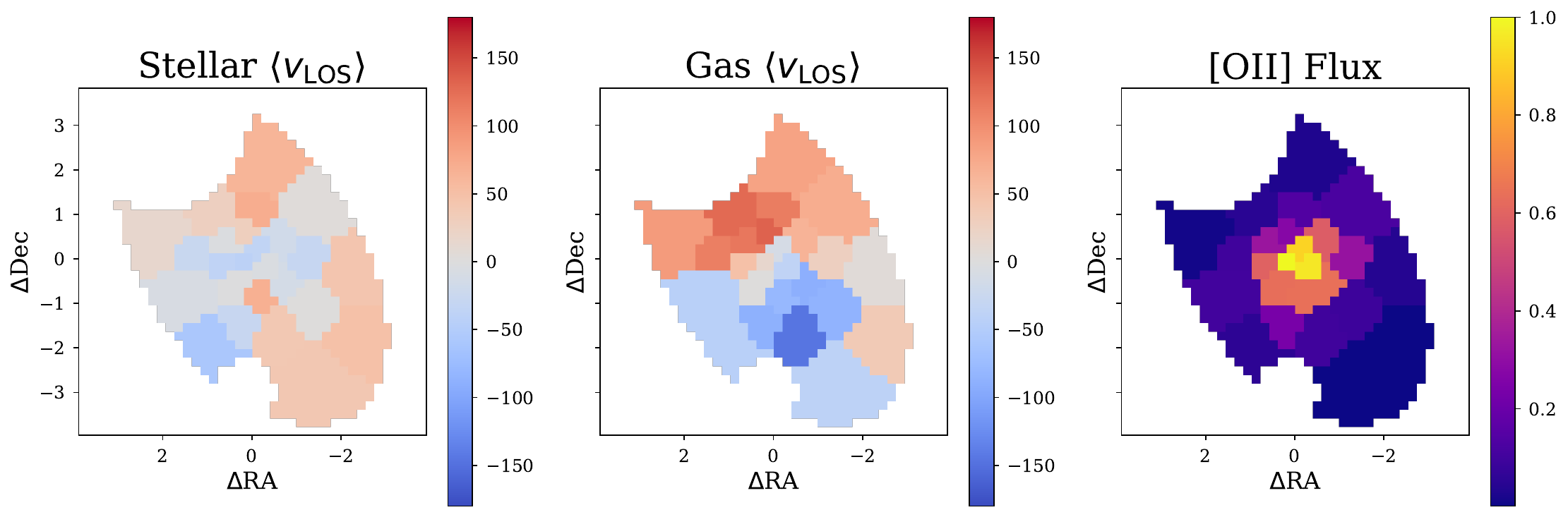}
    \caption{Additional features of the central galaxy of \MACSJ\, measured with MUSE IFU spectra. Axes show offsets from the BCG center in arcseconds. \textbf{Left:} The measured LOS velocity of the stellar component in km/s, showing negligible rotation. In all Voronoi bins, $(\langle v\rangle/\sigma_v)^2 < 0.04$. \textbf{Center:} The measured LOS velocity of the emission lines in km/s, showing clear evidence of rotation. \textbf{Right:} Observed flux-per-spaxel of [O II] 3729\AA, showing the spatial extent of gas emission. Flux is normalized by the brightest Voronoi bin.}
    \label{fig:gas_rotation}
\end{figure*}

We note that this central galaxy exhibits strong emission lines, potentially associated with an AGN and star formation. As such, we provide a list of emission line templates to \texttt{ppxf} in order to obtain the highest quality spectral fit. All emission lines are associated with a single kinematic component, separate from the stellar component. We include all emission lines in the relevant wavelength range known by \texttt{ppxf}, and manually add three lines not known by \texttt{ppxf}: [SII] 4069\AA, FII 5198\AA, and [NI] 5201\AA. While the kinematics of this gas component are not used to constrain our mass model, and have no bearing on the primary results of this paper, we note that this gas component is clearly kinematically decoupled from the stellar component, and shows clear evidence of rotation about the galactic center (see Fig.~\ref{fig:gas_rotation}). The stellar component shows negligible rotation, as expected in a BCG.

% I believe Cat no longer intends to study the BCG/AGN?
%An upcoming paper (Gibson et al., in prep) will present this potential AGN in more detail.

\subsubsection{\label{sec:stellar_kinematics_cov}Systematics Test and Covariance}

Measurements of stellar velocity dispersions can be highly sensitive to the wavelength range used, the choice of template library, and even the degree of the polynomial used to model background. As such, we conduct the following tests to verify the robustness of our kinematics map. Our approach follows \citet{Shajib.Mozumdar.ea2023}. We aim to bracket a range of reasonable settings, all of which compute high-quality kinematics maps, and we compute a full kinematics map at each setting as described in the section above. We choose four different `axes' along which we can reasonably change settings, and along each `axis', we choose between 2 and 6 settings. These four axes, and their various settings, are:

\begin{itemize}
    \item \textbf{Wavelength Range:} We extract and fit six different rest-frame wavelength ranges: 3700-4450\AA, 3750-4500\AA, 3800-4550\AA,
    3650-4550\AA, 3700-4600\AA, 3750-4650\AA.
    \item \textbf{Background Polynomial Order:} The degree of the additive polynomial used to represent the background. We use three separate values, 4 through 6.
    \item \textbf{Input Template Library:} We use the XSL stellar template library as described in Section~\ref{sec:stellar_kinematics_map}. As in \citet{Shajib.Mozumdar.ea2023}, we bisect the XSL library into two subsets, and compute kinematics maps with each subset. This results in three possible settings: `Full' XSL library, bisected half-`A', and bisected half-`B'.
    \item \textbf{Dust extinction:} The central galaxy of \MACSJ\, exhibits clear evidence of dust extinction, most evident in JWST NIRCam imaging (c.f. Fig~\ref{fig:position_constraints}, NE field). We utilize two possible settings. The first ignores any effect of dust on the spectral fits. The second uses the one-parameter attenuation curve presented in \citet{Calzetti.Armus.ea2000}, applying the same attenuation to both the stellar and gas components. The dust extinction is allowed to vary between 0 and 4 magnitudes, and is allowed a different value in each voronoi bin.
\end{itemize}

These constitute 108 possible settings, and produce 108 maps of the BCG's stellar kinematics. For each of these measured kinematics maps, we draw 1000 random realizations of the velocity dispersion map using the statistical errors produced by \texttt{ppxf}. From these 108,000 samples of the velocity dispersion in each voronoi bin, a mean and covariance are computed. The resulting best-fit kinematics map and correlation matrix are shown in Fig.~\ref{fig:kinematics-measurement}. This covariance $\boldsymbol{\Sigma}$ is used in a conventional multivariate-Gaussian likelihood, as described in Section \ref{sec:stellar_kinematics_jam}.

We find our resulting kinematics map to be consistent across all settings, with systematic uncertainty subdominant to statistical uncertainty in all but one spatial bin. Note that while \citet{Knabel.Mozumdar.ea2025} have demonstrated the leading systematic uncertainty comes from the choice of template library, and advocate comparing a wide array of such libraries, they demonstrate this uncertainty is at most $\mathcal{O}$(3-4\%) in fits to MUSE spectra. This is well below the $\approx 10\%$ uncertainty per bin measured in this work. See Appendix~\ref{appendix:kinematics_systematics} for further discussion.

\subsubsection{\label{sec:stellar_kinematics_mge}BCG Light Profile}

\begin{figure}
    %\centering
    \includegraphics[width=\linewidth]{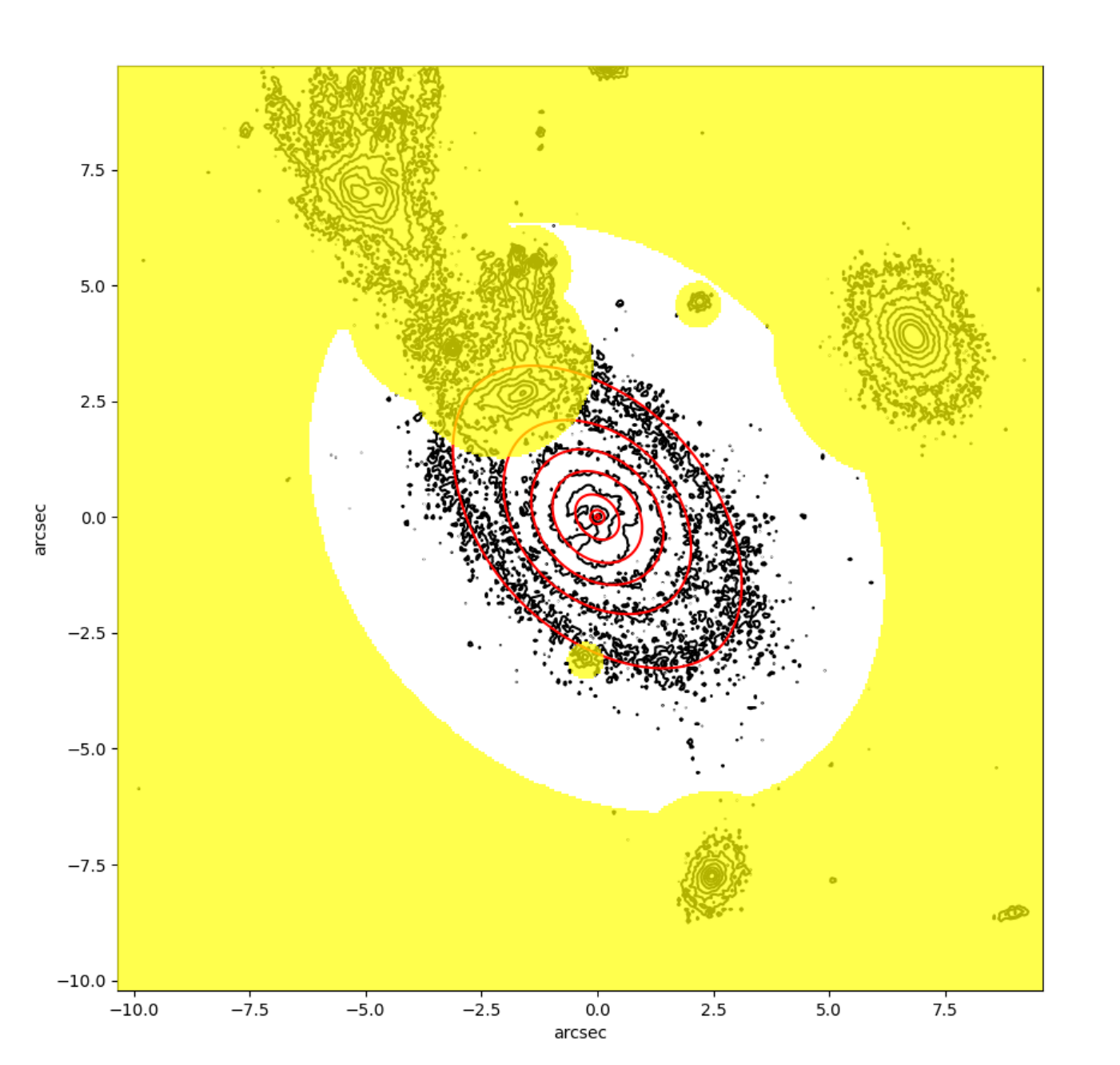}
    \caption{The MGE approximation of the central galaxy's light profile, as provided by \texttt{MGEFit}. Regions in yellow are excluded from the fit. Black contours represent ACS F555W imaging data, and red contours represent the MGE profile. See Section~\ref{sec:stellar_kinematics_mge} for details.}
    \label{fig:mgefit}
\end{figure}

The central galaxy's light profile is represented as a Multi-Gaussian Expansion (MGE), measured using \texttt{MGEFit}\cite{MGEFit} on the ACS/WFC F555W imaging. This band corresponds closely to the spectral wavelength range used to measure stellar kinematics. We apply \texttt{MGEFit} in an ellipse aligned with and centered on the BCG, with a semi-major axis of 7" and a semi-minor axis of 5.5", excluding several regions contaminated by member galaxies and other LOS structure, as shown in Figure~\ref{fig:mgefit}. Using an initial \texttt{ngauss} of 30, the resulting best-fit MGE contains 6 Gaussian components.

\subsubsection{\label{sec:stellar_kinematics_jam}JAM Model and Likelihood}

We compute the predicted stellar kinematics map using the widely used JAM package \cite{Cappellari2008,Cappellari2012,Cappellari2020}. JAM requires two primary inputs, each represented as an MGE: (1) the luminous tracer used to measure the kinematics (in this case, the light profile of the central galaxy), and (2) the mass profile in which
that tracer is embedded. The former is described in Section \ref{sec:stellar_kinematics_mge}. For the latter,
we combine the BCG stellar profile scaled by a mass-to-light ratio with the SIDM-modified
dark matter density profile. The dark density profile is decomposed into an MGE via the
\texttt{mge\_fit\_1d} function provided with \texttt{MGEFit}.

In addition, JAM requires several further parameters: a point spread function (PSF) model; the inclination angle relative to the line-of-sight; and the velocity anisotropy. The inclination angle, $i$, is also relevant to the triaxiality of the lensing halo, and has been described in Appendix \ref{appendix:triaxiality}. The meaning of the velocity anisotropy $\beta$ depends on whether the cylindrical \cite{Cappellari2008} or spherical \cite{Cappellari2020} alignment of the velocity ellipsoid is chosen, and whether the halo mass model is prolate or oblate. In this work, we use a spherical alignment of the velocity ellipsoid, where $\betasph = 1 - \sigma_\theta^2/\sigma_r^2$, and $\betasph = 0$ indicates isotropic velocity of the stellar population. As noted in \citet{Birrer.Buckley-Geer.ea2025}, JAM models with spherical and cylindrical alignment have been found to generally agree \cite{Cappellari2020,Zhu.Lu.ea2023}, and both \citet{Shajib.Mozumdar.ea2023,Birrer.Buckley-Geer.ea2025} use the spherical alignment in joint models of lensing and dynamics. Considerable uncertainty, however, exists on the value of the velocity anisotropy $\beta$; in this work, we adopt a flat prior on $\betasph \sim \mathcal{U}(-0.4, 0.4)$, slightly wider than the allowed values in \cite{Shajib.Mozumdar.ea2023,Birrer.Buckley-Geer.ea2025}. %{\color{red}Radial variation in the anisotropy? We don't allow it: Anowar's paper tests it and finds negligible difference. However, that's not a cluster. Sagunski does not consider it either, but they do allow radial gradient in $\Upsilon_*$.}

We use a simple Gaussian model of the PSF of our IFU data, from which the kinematics are measured. Metadata provided with the MUSE cube suggest the seeing varied between FWHM 0.66" and 0.8" over the course of the observations. In addition, we perform a simple test to compare our the relative seeing between the HST F555W imaging and the MUSE IFU data. Given the MGE model computed in Section \ref{sec:stellar_kinematics_mge}, we convolve by a single-Gaussian PSF model to compute a predicted, PSF-smoothed image, and compare to the IFU cube summed between 5300 and 5700\AA. Residuals between these suggest the true PSF at the MUSE slicer could be higher, up to around $\sim1.15$" FWHM. We thus assume a single-component Gaussian PSF, and adopt a prior on its size $\sigma_{\text{PSF}} \sim \mathcal{U}(0.35",0.5")$, spanning FWHM values of approximately 0.5" to 1.2".

Given all the above inputs, for each spatial bin $i$, the predicted luminosity-weighted velocity dispersion is:

% c.f. Shajib 2023 Section 4.2 & 4.4
\begin{equation}
    \langle\sigma_{\text{los},i}^2\rangle = \frac{\int_{\Theta_i}  I \langle v^2_{\text{los}}\rangle \circledast \text{PSF}\, d^2\theta}{\int_{\Theta_i}  I \circledast \text{PSF}\, d^2\theta} 
\end{equation}

Where $\Theta_i$ denotes the aperture of the $i$th spatial bin, $I$ represents the projected surface brightness of the tracer population, $\circledast \text{PSF}$ denotes convolution by the PSF, and $\langle v^2_{\text{los}}\rangle$ is the predicted LOS velocity dispersion at position $\theta$ computed by JAM. As in \citet{Shajib.Mozumdar.ea2023}, we take the first velocity moment to be negligible, leaving $\langle\sigma_{\text{los}}^2\rangle = \langle v_{\text{los}}^2\rangle$. (See also Figure \ref{fig:gas_rotation}.) With our measured kinematics map $\boldsymbol{\sigma^{\text{obs}}_{\text{los}}} =\{\sigma_{\text{los},i}^{\text{obs}}\}$ and covariance matrix $\boldsymbol{\Sigma}$, the likelihood is therefore:

\begin{align}
    \Delta \boldsymbol{\sigma} &= \boldsymbol{\sigma^{\text{obs}}_{\text{los}}} - \langle \boldsymbol{\sigma_{\text{los}}}\rangle \\
    \mathcal{L}(\boldsymbol{\sigma^{\text{obs}}_{\text{los}}} | \mathcal{M}) &= \frac{1}{\sqrt{(2\pi)^{k} |\boldsymbol{\Sigma}|}}
    \exp\big(-\frac{1}{2} (\Delta \boldsymbol{\sigma})^T \boldsymbol{\Sigma}^{-1} (\Delta \boldsymbol{\sigma})\big)
\end{align}

Where $\mathcal{M}$ denotes our model parameters.

\section{\label{sec:results}Results}

\begin{table*}
    \input{results_table}

    \caption{\textbf{Results:} Priors and posteriors for all three models fit in this work. In the priors column, $\mathcal{U}(a,b)$ denotes a uniform prior on the interval $[a,b]$, and $\mathcal{N}(\mu,\sigma)$ denotes a Gaussian prior of mean $\mu$ and variance $\sigma^2$. The `Result' columns list the median and $1\sigma$ bounds on each parameter. The shared, lensing-only, and kinematics-only parameters and priors are explained in detail in Sections~\ref{sec:methods_sidm_model}, \ref{sec:methods_strong_lensing}, and \ref{sec:methods_stellar_kinematics} respectively; for convenience, brief descriptions of each parameter are included here.}
    \label{tab:results}
\end{table*}

Given the model, constraints, and likelihoods described in Section \ref{sec:methods}, we compute three sets of posteriors. First, we present `lensing-only' results, using only the strong lensing constraints; second, we present the `kinematics-only' constraints, derived only from the measured kinematics map of the BCG; lastly, we present a joint model of the combined lensing and kinematics constraints.

All cosmology-dependent quantities use the best-fit parameters from Planck 2015 (P15) \cite{Planck15}. The following posteriors were obtained with the \texttt{emcee} ensemble MCMC sampler 
\cite{emcee}. For the kinematics-only constraints, 80 walkers were used, whereas the lensing-only and joint results utilized 48 walkers each. While the kinematics-only result used a single sampler run, the lensing-only and joint posteriors were each estimated with two independent sampler runs in order to check the consistency of results. All samplers used a combination of `moves', primarily the original \texttt{StretchMove} presented in \citet{Goodman.Weare2010}, but with some additional weight given to \texttt{DEMove} and \texttt{DESnookerMove} \cite{Nelson.Ford.ea2014,TerBraak2006} to better explore multimodal posteriors.

All corner plots in this work were made with \texttt{corner} \cite{corner}. In all corner plots of posteriors, 2D contours shade the 1- and 2-$\sigma$ posteriors, and 1D posteriors overlay the $-1\sigma$, median, and $+1\sigma$ confidence intervals with dotted vertical lines. The priors and posteriors of all parameters for all three models are shown in Table~\ref{tab:results}.

\subsection{\label{sec:results_lensing_only}Lensing-Only Results}

\begin{figure}
    \centering
    \includegraphics[width=1\linewidth]{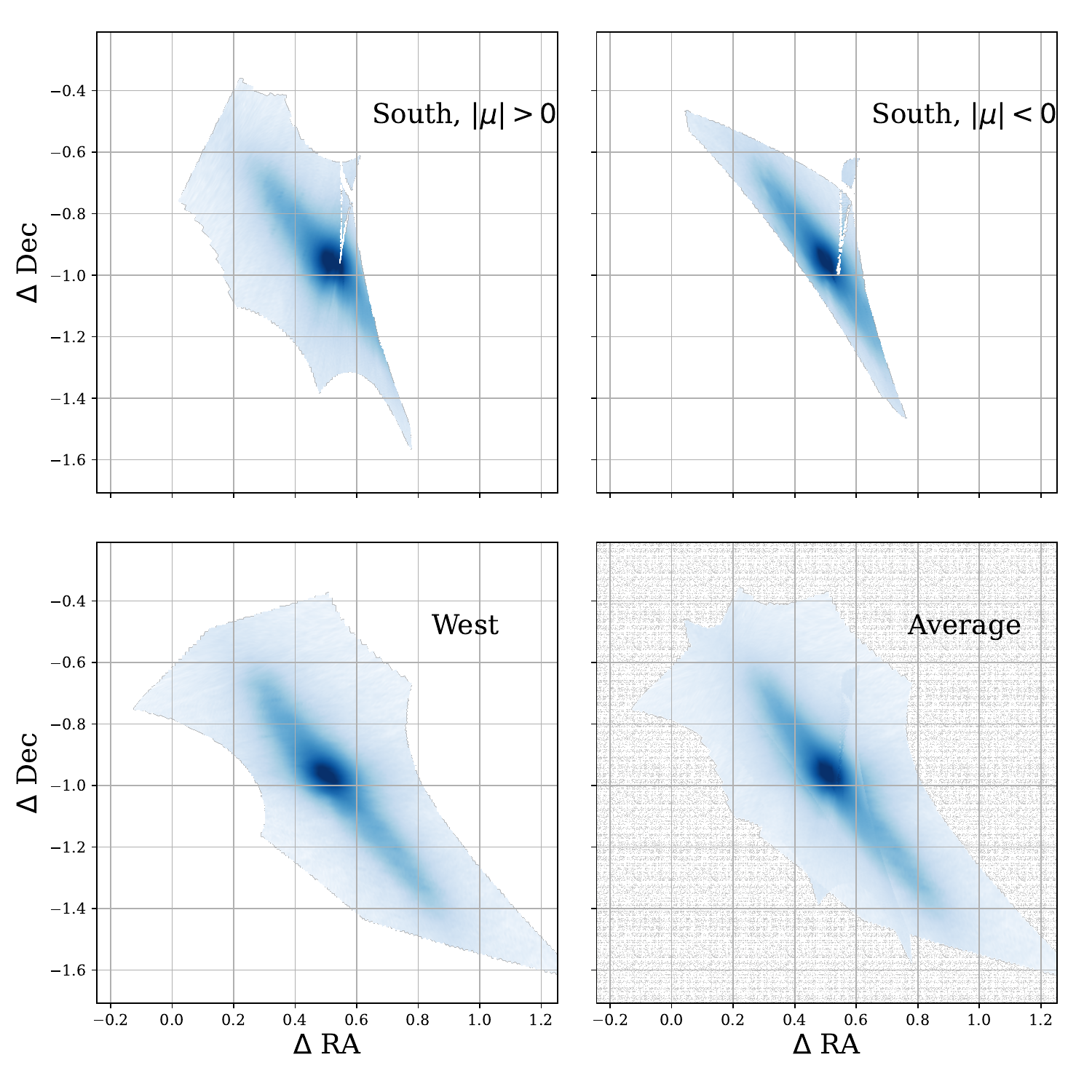}
    \caption{\textbf{Lensing-Only Fit:} Source plane projection of the observed lensing arcs of the primary source at z$\approx$1.95. \textbf{Top:} Luminosity of the southern arc, separated into the positive and negative magnification images. \textbf{Bottom Left:} The luminosity of the western arc. \textbf{Bottom Right:} The average of the former three images, demonstrating their consistency.}
    \label{fig:lensing-only-fit}
\end{figure}

The lensing-only model consists of seventeen parameters, constrained by 29 measured positions of 10 source features listed in Table~\ref{tab:position_constraints}. As stated above, two independent \texttt{emcee} samplers of 48 walkers each were run for this model, in order to verify the robustness of posteriors. Both samplers ran for over 1,220,000 iterations, and yielded consistent estimates of the mean autocorrelation time of $\sim52,450$ and $\sim52,270$ iterations. This extremely long autocorrelation time suggests that \texttt{emcee} struggled to efficiently sample this parameter space, likely due to a complex likelihood surface. Nevertheless, we consider these results to be robust: the two independent samplers produced posteriors which are quantitatively and qualitatively consistent.

The predicted source-plane light derived from the southern and western arcs is shown in Figure~\ref{fig:lensing-only-fit}, using the best fit mass model and NIRCam F150W imaging. The imaging was cropped to regions tightly bracketing the observed arcs, and the positions of each observed pixel were projected to the source plane given the best-fitting lens model. They were then interpolated to a grid of source-plane pixels with the scipy method \texttt{griddata}.

The lensing favors an overall DM halo with ellipticity closely aligned to the axis of the central galaxy, yielding $\phi = 44.3^{+1.8}_{-
 1.7}$ degrees East-of-North, compared to the BCG's position angle of $42\degree$. The position of the overall DM halo is also consistent with the center of the BCG, although the lensing alone favors a DM halo centered slightly to the south.

 The lensing only model gives constraints on the mass-to-light ratio, mass, and concentration of $\log_{10}\Upsilon_* = 0.5^{+0.1}_{-0.2}$, $\log_{10}(M_{\text{200m}}/M_{\odot}) = 14.9 \pm 0.2$, $c_{\text{200m}} = 10^{+2}_{-3}$.  The isothermal radius is constrained to be $\log_{10}(r_1/\text{kpc}) = 2.3^{+0.3}_{-0.9}$.

% \reviewer{There is a kinematics residual in Fig. 8. Could the authors also show the
%residuals for the lens model?
%}

\subsection{\label{sec:results_kinematics_only}Kinematics-Only Results}

\begin{figure}
    \centering
    \includegraphics[trim={0 1.7cm 1cm 2cm},clip,width=1\linewidth]{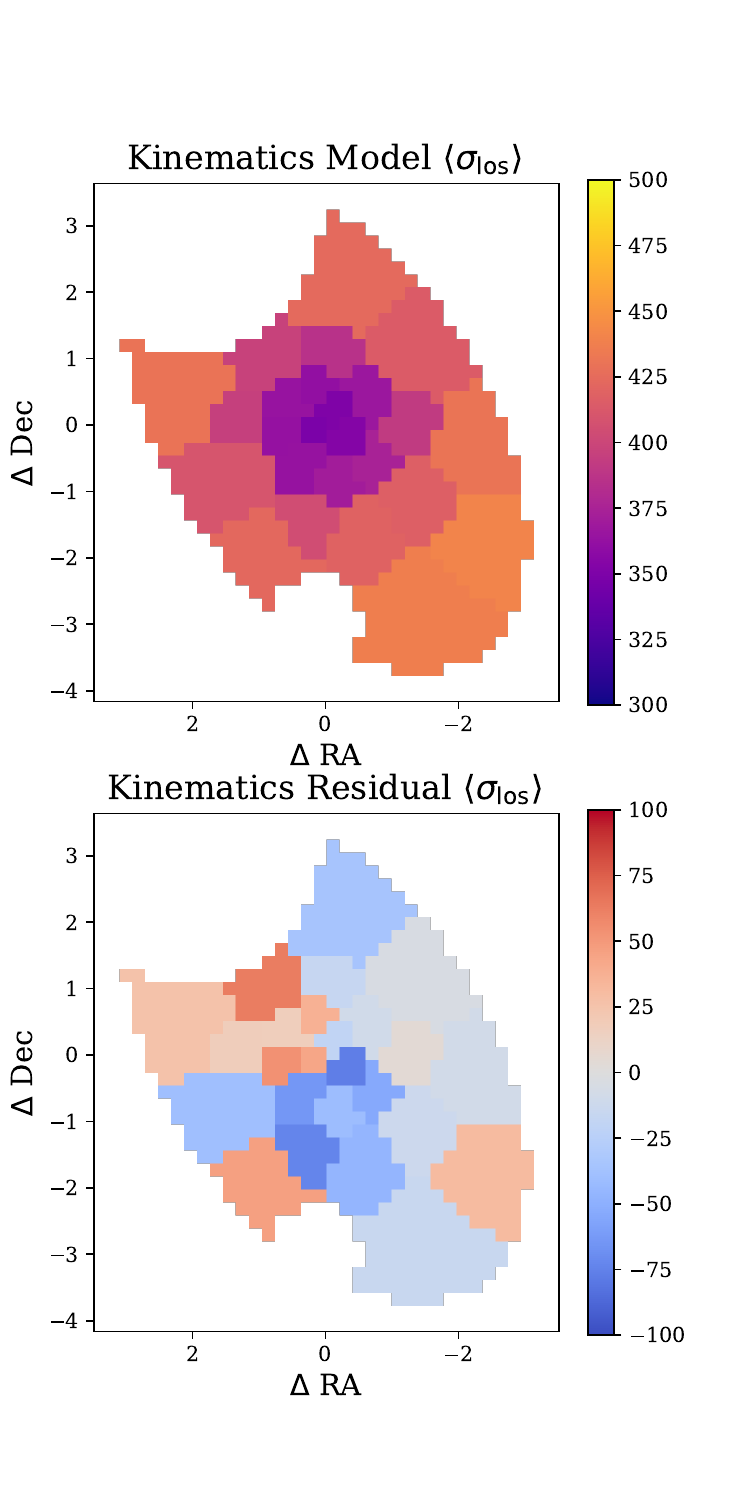}
    \caption{\textbf{Kinematics Fit:} The best-fit predicted kinematics map and residual (predicted - data). Axes units are arcsec relative to BCG center, color map units are km/s.}
    \label{fig:kinematics-best-fit}
\end{figure}

The kinematics model has nine parameters (the seven shared parameters describing the BCG and cluster DM halo, and $\betasph$ and $\sigma_{\text{PSF}}$), and is constrained by the measured velocity dispersion in 25 Voronoi bins. The 80-walker \texttt{emcee} chain for this model ran for 192,648 iterations, with a mean autocorrelation time of 910 iterations across all nine parameters. Having run for over 200 autocorrelation times, we consider these samples to be exceedingly stable and well converged.

This model provides constraints of $\log_{10}\Upsilon_* = 0.4^{+0.2}_{-0.3}$, $\log_{10}(M_{\text{200m}}/M_{\odot}) = 14.5 \pm 0.6$, $c_{\text{200m}} = 11 \pm 3$, and $\log_{10}(r_1/\text{kpc}) = 1.7^{+0.6}_{-1.0}$. These are fully consistent with the lensing-only model. %As expected, the kinematics alone provide very weak constraints on the mass and concentration of the overall halo in compared to strong lensing.

The predicted kinematics map of the maximum likelihood sample is shown in Figure~\ref{fig:kinematics-best-fit}. The best-fit model yields a reduced $\chi^2$ of 2.76. As expected, the stellar kinematics is most sensitive to the total mass within $\sim20$kpc, and has only weak constraining power on the mass and concentration of the overall halo. It is reassuring, however, that the kinematics model alone yields parameters that are highly consistent with the lensing-only model: the posteriors on $\Upsilon_*$ are similar, and in $M_{\text{200m}}$-$c_{\text{200m}}$ space, the $1\sigma$ posteriors of the lensing-only model lie almost entirely within those of the kinematics-only model. A larger difference can be seen in the SIDM cross section $\sigma/m$, where the lensing alone prefers markedly higher values. Nevertheless, they are consistent, justifying a combination of both lensing and kinematics constraints.
% JOD: Shajib 2025 (TDCOSMO XXIII) uses a likelihood ratio test to check if it's valid to combine SL and kinematics. We could do something similar with SL and kinematics here..?

Our kinematics model prefers an anisotropy $\betasph < 0$, implying $\sigma_{\theta}/\sigma_{r} > 1$. This preference only strengthens when lensing constraints are included, see Table~\ref{tab:results}. \citet{Shajib.Treu.ea2025} similarly found $\sigma_{\theta}/\sigma_r > 1$ in both ground- and space-based IFU measurements of a galaxy-scale lens. While the PSF size $\sigma_{\text{PSF}}$ is entirely unconstrained, it fortunately demonstrates almost no degeneracy with any other parameters.

\subsection{\label{sec:results_combined}Combined Strong Lensing and Kinematics Results}

\begin{figure}
    \centering
    \includegraphics[width=1\linewidth]{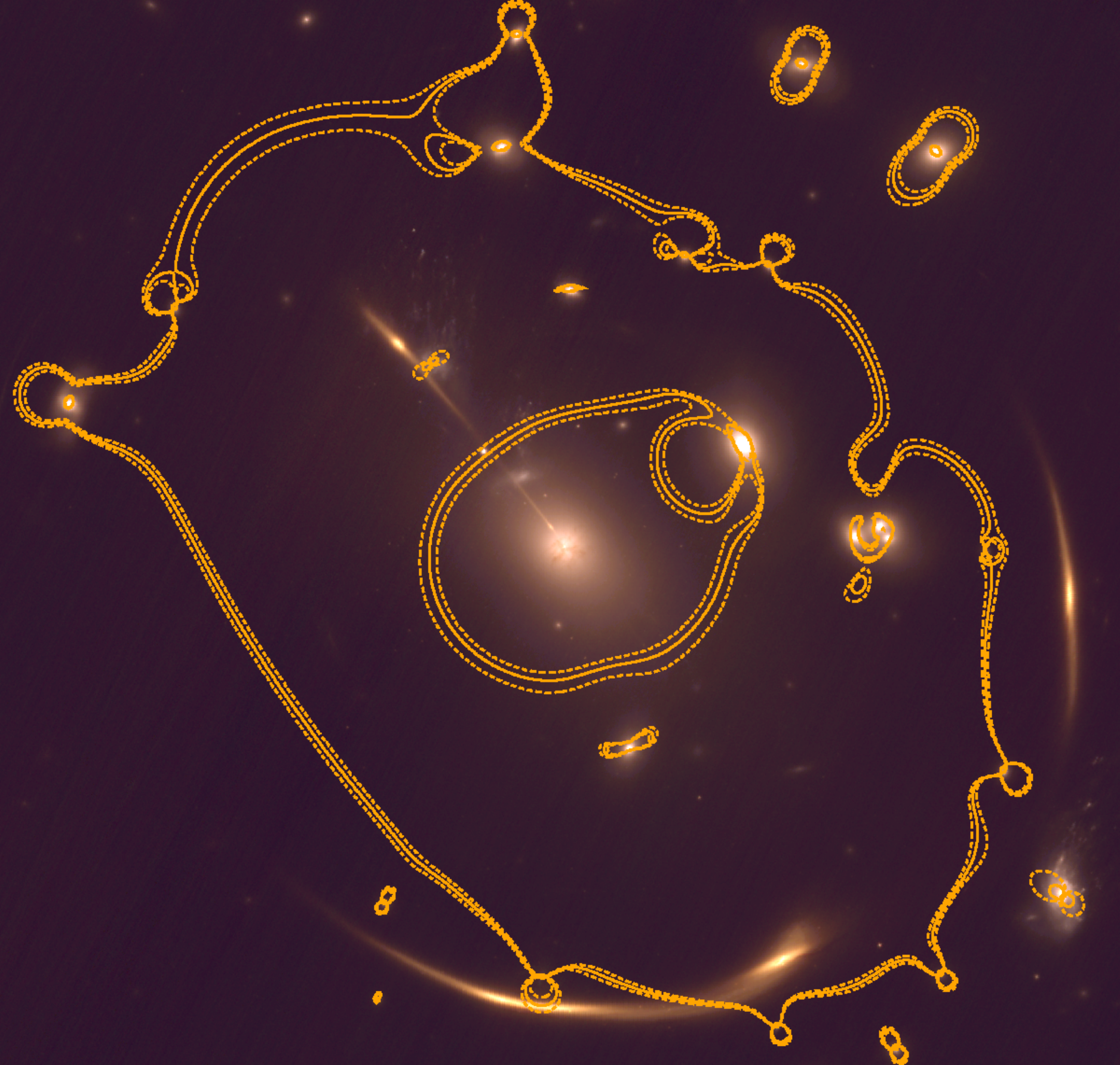}
    \caption{\textbf{Critical Curves:} Posteriors on the lensing critical curves of the joint model. Median critical curves are displayed as a solid line, $\pm1\sigma$ critical curves are dashed lines. Color bands and scaling are identical to Figure~\ref{fig:members}.}
    \label{fig:critical-curves-joint}
\end{figure}

\begin{figure*}
    \centering
    \includegraphics[width=1.0\linewidth]{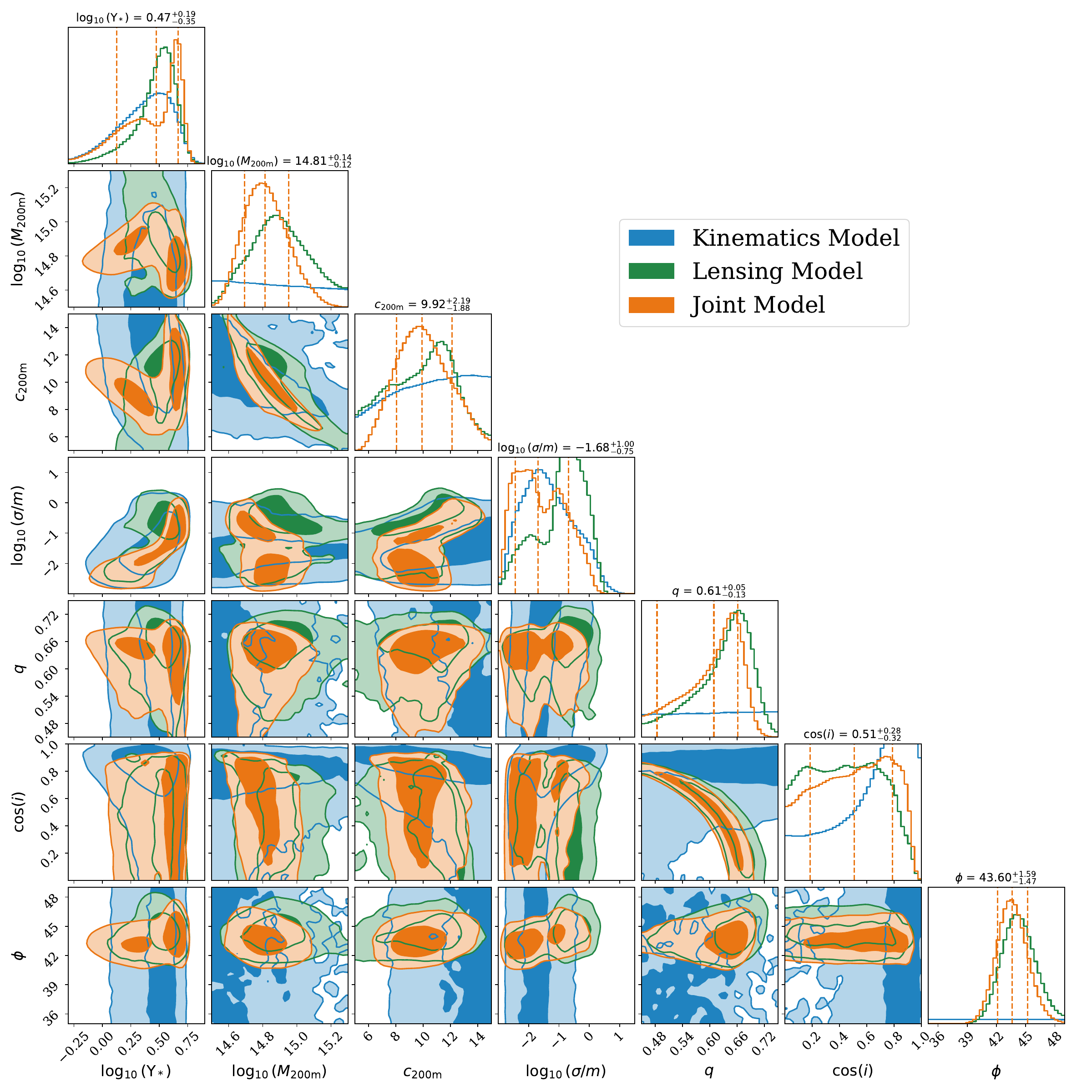}
    \caption{Posteriors of shared parameters across all models. For every pair of parameters, colored contours show the 1- and 2-$\sigma$ posteriors. At the top of each column, the median and $\pm1\sigma$ posteriors are listed for the joint model, and marked with dashed vertical lines.}
    \label{fig:corner-all-models}
\end{figure*}

The results of the joint model, combining strong lensing and stellar kinematics, are the primary result of this work. 
As in Section~\ref{sec:results_lensing_only}, we run two separate \texttt{emcee} samplers of 48 walkers each to verify the robustness of these results. Reassuringly, \texttt{emcee} was able to sample the joint model more efficiently than the lensing-only model, despite having slightly higher dimensionality and a more expensive likelihood. This suggests the kinematics constraints contribute to a `smoothing-out' of the likelihood surface. The two joint samplers both ran for over 660,000 iterations, yielding a mean autocorrelation time across all parameters of $\sim21,400$ and $\sim22,900$ iterations, substantially shorter than for the lensing-only model. Both independent sampling runs yielded nearly identical posteriors, and we consider these posteriors to be robust.

The predicted critical curves (positions of infinite magnification) from this model are displayed in Figure~\ref{fig:critical-curves-joint}. To create this plot, the magnification field $\mu(\vec\theta)$ was computed from 10,000 randomly drawn samples of the posterior. The median ($\pm1\sigma$) critical curves are the contours where the median ($15.9\%, 84.1\%$ quantile) of $\mu^{-1}$ is zero. As expected, the tangential critical curve divides the southern arc neatly into two images of positive and negative magnification, and the radial critical curve divides the central and North-East images near a point where their light is dimmest.
These curves are very consistent with those derived from the lensing-only model, which is not overlaid to improve the clarity of the plot. The predicted magnification at the conjugate points, however, differs between the joint and lensing-only models; see Section~\ref{sec:supernovae} for further discussion. The best-fit kinematics map from this joint model is nearly identical to that of the kinematics-only model, with a (non-reduced) $\chi^2$ of 45.5, compared to 44.3 for the kinematics model shown in Figure~\ref{fig:kinematics-best-fit}.

This joint model yields constraints of $\log_{10} \Upsilon_* = 0.5^{+0.2}_{-0.4}$, $\log_{10}(M_{\text{200m}}/M_{\odot}) = 14.8 \pm0.1$, $c_{\text{200m}} = 10 \pm 2$, and $\log_{10}(r_1/\text{kpc}) = 1.6^{+0.7}_{-1.0}$. For numerical constraints on other parameters, see Table~\ref{tab:results}. Curiously, the joint posteriors exhibit a strong bimodality absent from either the kinematics- or lensing-only models. This is most evident in the $\Upsilon_*$-$M_{\text{200m}}$ space, see Fig.~\ref{fig:corner-all-models}. 
%\reviewer{ the authors mentioned that there
%is a strong degeneracy between the stellar mass-to-light ratio and the cross
%section. The authors need to explain the underlying physics causing this
%Degeneracy.}
%\jackcomment{We do discuss this in the Discussion section}
In addition, the stellar mass-to-light ratio $\Upsilon_*$ is strongly degenerate with the SIDM cross section $\sigma/m$, with higher cross sections occurring exclusively at high $\Upsilon_*$. 

\begin{figure}
    \centering
    \includegraphics[trim={0.5cm 0 1cm 1cm},clip,width=1\linewidth]{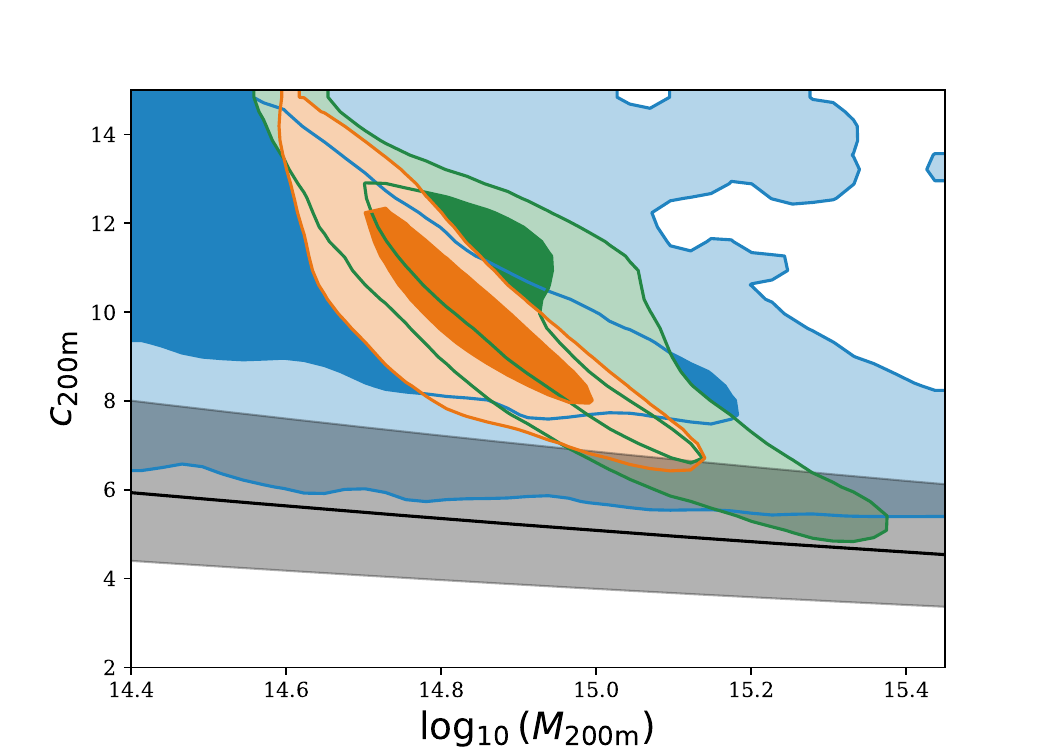}
    \caption{\textbf{Mass-Concentration:} A comparison of the posteriors on $M_{\text{200m}}$ and $c_{\text{200m}}$ in this work to theoretical predictions. The mass-concentration relation of L16 is displayed in black, with a shaded band showing 0.13 dex scatter. The L16 relation was computed at the cluster redshift, and translated to $M_{\text{200m}}, c_{\text{200m}}$ using Colossus \cite{ColossusDiemer}.}
    \label{fig:mass-concentration}
\end{figure}

While the kinematics and lensing models alone both prefer a cluster halo with relatively high concentration, the significance of this preference increases in the joint model. The posteriors of all models are reproduced in Figure~\ref{fig:mass-concentration}, and compared to the expected concentration from \citet{Ludlow.Bose.ea2016a} (L16) with a scatter of 0.13 dex. These choices are identical that used in \citet{Robertson.Massey.ea2021}. The joint model prefers a halo concentration around $\sim2\sigma$ higher than the L16 relation.
%\reviewer{The results presented in Fig. 11 and Fig. 15 show certain discrepancies with the predictions and the existing measurements. Discussions about these differences would be very helpful. }
Strong lensing has a strong selection function for halos that are massive, elongated along the line-of-sight, or, indeed, highly concentrated, so this result is not surprising. See Section~\ref{sec:discussion} for further discussion.
% JOD: What can we compare to here? Are there literature studies of concentration of lensing clusters?
\subsection{\label{sec:results_external_shear}External Shear}

The lensing-only and joint models above yielded external shear with magnitude $|\gammaext|$ of $0.09 \pm 0.02$ and $0.07 \pm 0.01$, respectively. The external shear term is intended to represent the contribution of cosmic shear sourced by large scale structure, and other mass near the line of sight, which are fundamental systematic uncertainties in strong gravitational lensing. While the motivation of this approach is to generically capture and account for any such perturbations, \citet{Etherington.Nightingale.ea2024} has demonstrated in galaxy-galaxy lenses that the magnitude and direction of SL-derived $\gammaext$ shows little correlation with known weak lensing shear, and is instead highly correlated with the elliptical orientation of the lens itself. They suggest that, in practice, the external shear accounts for structure of the lens mass which parametric models fail to properly describe. 

%JOD: External Shear Adds Uncertainty to time delay {\color{red} Furthermore, the external shear can add uncertainty to time delay predictions... (JOD find out if others have mentioned this)}

To investigate whether the measured $\gammaext$ may be explained physically, we conduct the following test. In the \MACSJ\, field, there is a luminous red galaxy $\sim2.5'$ to the south, which is likely associated with the \MACSJ\, cluster and may host a significant DM halo of its own. We conduct another \texttt{emcee} sampling of the joint likelihood, including a further NFW halo of fixed $c_{200m}=6$ and varied $M_{\text{200m}}=M_c$ at the position of this bright LRG, (RA, Dec) =(01:38:06, -21:57:51). To simplify the parameter space, we identify six parameters with negligible degeneracy, and fix them to the median values obtained by the joint sampling in Section~\ref{sec:results_combined}. These parameters are $\alpha$, $r_s^{\text{J4}}$, $\sigmalt^{\text{J1,J3,J4}}$, $\sigma_{\text{PSF}}$.

\begin{figure}
    \centering
    \includegraphics[width=1.0\linewidth]{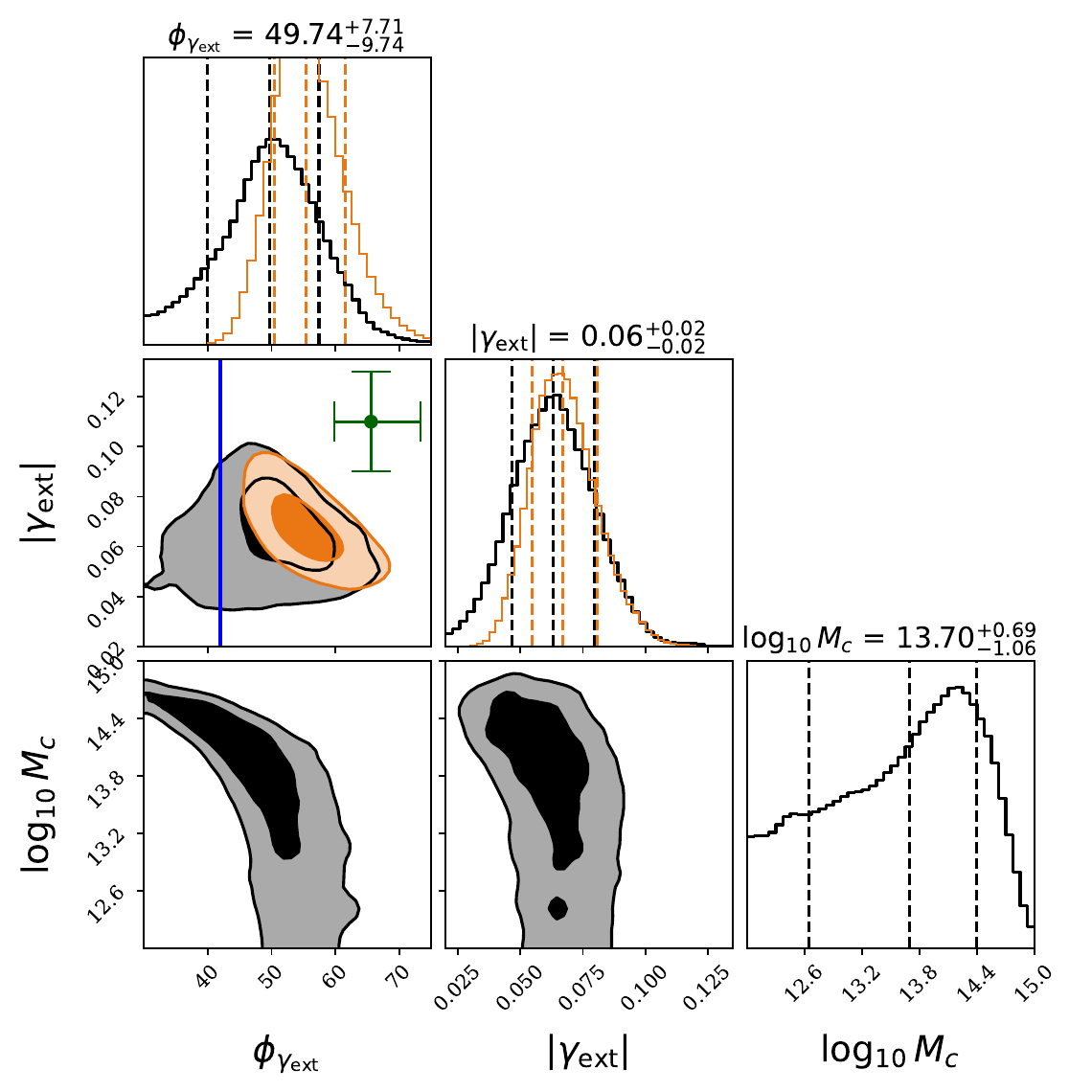}
    \caption{
    \textbf{External Shear:} The direction and magnitude of the external shear $\gammaext$. Black contours show the results of the test described in Section~\ref{sec:results_external_shear}, and orange contours denote the results of the joint model. In $\phi_{\gammaext}$-$|\gammaext|$ space, a blue line marks the orientation of the central galaxy of \MACSJ, and a green point marks the result of \cite{Acebron.Bergamini.ea2025}.}
    \label{fig:gamma_ext_corner}
\end{figure}

The results of this test are illustrated in Figure~\ref{fig:gamma_ext_corner}, showing posteriors on the position angle and magnitude of $\gammaext$, as well as the mass of the nearby NFW halo, $M_c$. \citet{Acebron.Bergamini.ea2025} independently conducted a similar test, sampling separate models with a mass component associated with the same luminous galaxy to the south. The $\gammaext$ measured by their \texttt{reference} model is overlaid for comparison. Allowing for the nearby mass slightly decreases the derived shear magnitude, suggesting this mass can only partly explain the observed shear. The data clearly prefer a substantial mass at the location of the LRG, with a mass up to $\sim10^{14.4}$ $M_{\odot}$. While the effect on the shear magnitude $|\gammaext|$ is slight, the nearby mass has a significant effect on the position angle of the shear, extending the posteriors to include the orientation of the central galaxy of \MACSJ. Similarly, \citet{Etherington.Nightingale.ea2024} found that $\gammaext$ frequently aligned with the lens orientation. Our lensing constraints additionally show a clear preference for the presence of this other mass: this test yielded a minimum lensing $\chi^2$ of 3.5, significantly lower than the best lensing $\chi^2$ of 4.08 produced by the fiducial joint model. Nevertheless, the presence of the extra mass results in little change to our parameters of interest compared to the joint model; the posteriors on $\Upsilon_*$, $M_{\text{200m}}$, $c_{\text{200m}}$, and $\sigma/m$ are all consistent within $1\sigma$.

% Unfortunately I don't currently have the time to do this correctly
%\subsection{\label{sec:time_delays}Implications on Future Observations of SNe Requiem and Encore}
%{\color{red} TODO}

\subsection{\label{sec:supernovae}Supernovae Requiem and Encore}

\begin{figure*}
    \centering
    \includegraphics[trim={5cm 0 5cm 0},clip,width=1.\linewidth]{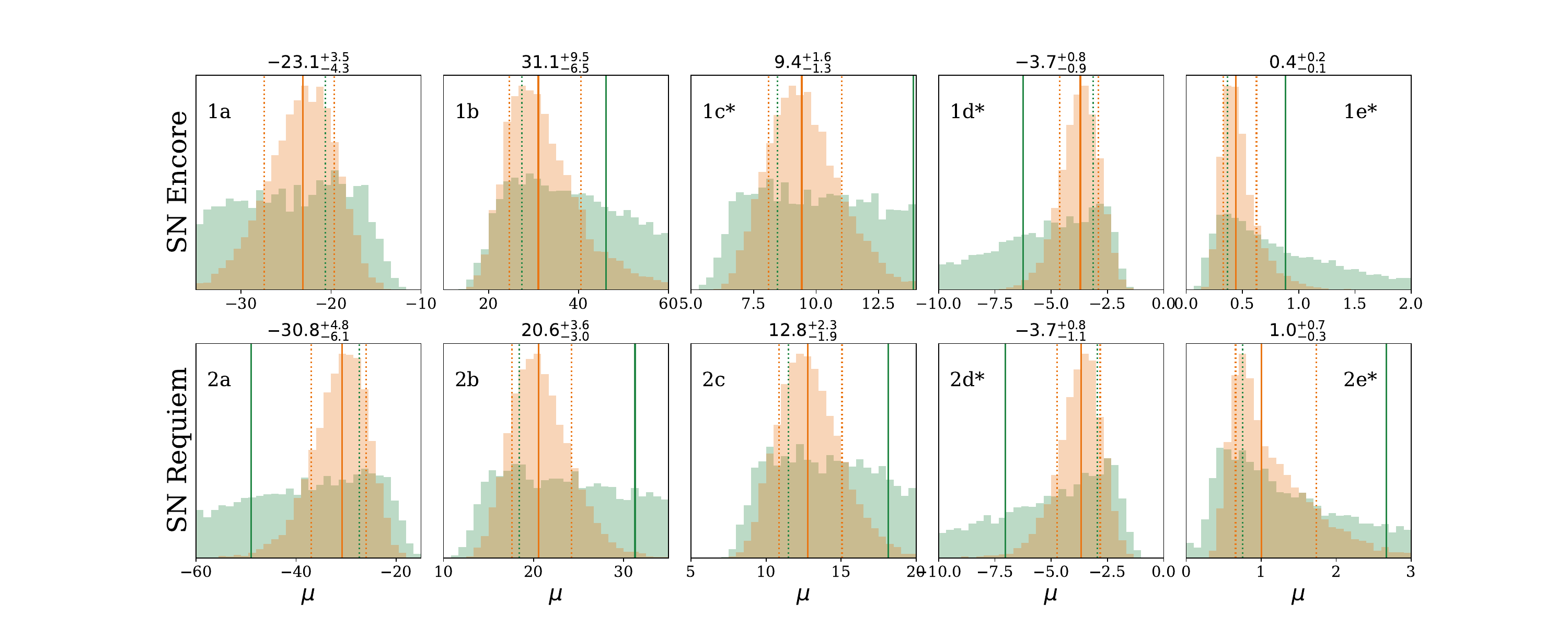}
    \caption{\textbf{Model SNe Magnifications:} The magnification of the observed and predicted images of SNe Requiem and Encore. Each histogram is generated from 10,000 randomly chosen samples of the \texttt{emcee} chains. The posteriors of the lensing-only and joint model are shown in green and orange respectively, and their median ($1\sigma$) limits are overlaid as solid (dotted) vertical lines. The title of each subplot lists the result of the joint model. Images which are not used to constrain the model are marked with an asterisk (*), including image groups (d*) and (e*) which have not been observed.}
    \label{fig:sne-magnifications}
\end{figure*}

Here we present model-predicted time delays and magnifications of both lensed SNe.  The predicted magnifications of previously observed SNe images are shown in Figure~\ref{fig:sne-magnifications}. For clarity, we use the same labels as \citet{Acebron.Bergamini.ea2025,Ertl.Suyu.ea2025}, and mark images which are not used as model constraints with an asterisk (*). In this nomenclature, images (2a, 2b, 2c) and (1a, 1b) correspond respectively to (SN-R.2, SN-R.1, SN-R.3) and (SN-E.2, SN-E.1) in our Table~\ref{tab:position_constraints}. We additionally include predicted images (d*) and (e*), with predicted positions in the central and North-East radial arcs where neither supernova has yet been observed. While the lensing-only model provides very poor constraints on the magnifications, our joint model including kinematics yields clear predictions.

Time delays predicted by the joint model are listed in Table~\ref{tab:time-delays}. These are computed using the fixed P15 cosmology, with H$_0 = 67.7$ km s$^{-1}$ Mpc$^{-1}$ and a time-delay distance $D_{\Delta t}$ of $\sim1860$ Mpc for this lens and source. As in \citet{Rodney.Brammer.ea2021}, time delays are normalized relative to the southernmost image of both SNe, labeled (b) here. Dates of peak brightness are estimated from the observed SN ages reported by \citet{Rodney.Brammer.ea2021} and \citet{Dhawan.Pierel.ea2024b} for Requiem and Encore, respectively. These results imply the next observed image will be feature (d) of SN Requiem in late 2027 with $\sim1$ year uncertainty, approximately two years from the date of this publication.

\begin{table}[]
    \renewcommand\arraystretch{1.15}
    \centering
    \begin{tabular}{c| >{\centering\arraybackslash}p{1.7cm} |>{\centering\arraybackslash}p{1.8cm}|>{\centering\arraybackslash}p{1.7cm}|>{\centering\arraybackslash}p{1.8cm}}
        Feature & \multicolumn{2}{c|}{Requiem (2)} & \multicolumn{2}{c}{Encore (1)} \\
        Group & Delay (d) & Median  & Delay (d) & Median \\
         \hline
        a & $55 \pm 6$ & Jun 2016 & $41 \pm 4 $ & Sept 2023 \\
        b & $0$ & Apr 2016 & $0$ & July 2023 \\ 
        c & $-31 \pm 5$ & Mar 2016 & $-300^{+31}_{-34}$ & Sept 2022 \\
        d & $4245^{+343}_{-322}$ & \textbf{Nov 2027} & $3507^{+291}_{-280}$ & \textbf{Feb 2033} \\
        e & $4372^{+339}_{-323}$ & \textbf{Apr 2028} & $3819^{+307}_{-283 }$ & \textbf{Jan 2034} \\
    \end{tabular}
    \caption{\textbf{Time Delays:} The predicted time delays between all five images of both lensed SNe, given in days. Delays are normalized relative to the southernmost image (b). The `Median' columns give the median date of predicted peak SN brightness, rounded to the month. Image groups (d) and (e) are anticipated in the next decade; their dates are shown in bold.}
    \label{tab:time-delays}
\end{table}

\section{\label{sec:discussion}Discussion}

\begin{figure*}
    \centering
    \includegraphics[trim={3cm 0 3cm 0.5cm},clip,width=1\linewidth]{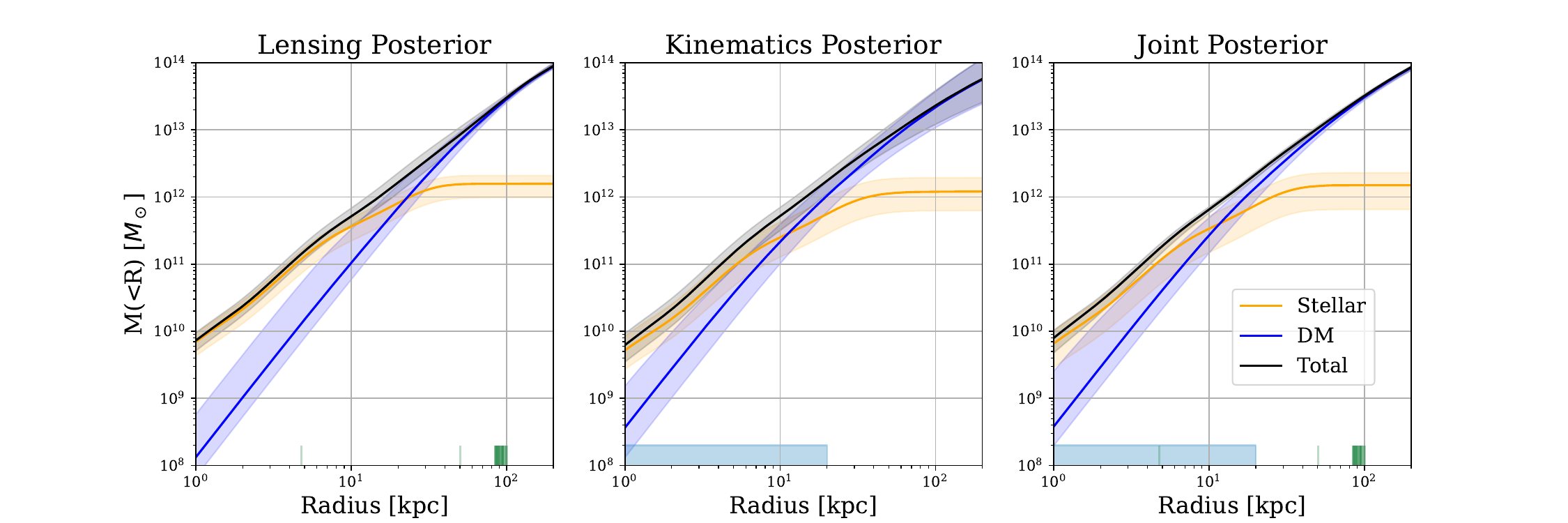}
    \caption{\textbf{Enclosed Mass:} Posteriors on the enclosed 3D mass for all three models presented here. At the bottom of each plot, a shaded blue region represents the radii constrained by stellar kinematics, and vertical green lines represent the projected radii of strong lensing constraints.}
    \label{fig:enclosed-mass}
\end{figure*}

We have presented independent and joint constraints from strong lensing and stellar kinematics models. The constraints on parameters of interest (chiefly the stellar mass-to-light ratio $\Upsilon_*$, halo mass and concentration, and SIDM cross-section) are remarkably consistent between the individual probes. The joint model provides notably tighter constraints on the mass and concentration, and a more stringent upper limit on the SIDM interaction cross-section. As a means of comparison between all three models in this work, the enclosed 3D mass relation $M(<r)$ for the DM, BCG stellar, and total mass are displayed in Figure~\ref{fig:enclosed-mass}.

 Notably, our inferred NFW concentration on the cluster dark matter halo is $\sim2\sigma$ higher than the predicted cosmological mass-concentration relation (see Fig.~\ref{fig:mass-concentration}). However, this does not represent a significant tension with conventional cosmological theory. It is widely known that strong lensing-selected halos are biased to high concentrations, see for example \citet{Meneghetti.Rasia.ea2014}. Additionally, dynamically relaxed halos are expected to have higher concentrations \cite{Meneghetti.Rasia.ea2014,Child.Habib.ea2018,Darragh-Ford.Mantz.ea2023}; our result on $c_{\text{200m}}$ may be further evidence of \MACSJ's relaxed state.

Our results exhibit a strong bimodality in the stellar mass-to-light ratio $\Upsilon_*$, and less so in the SIDM cross section. These modes correspond to a strong degeneracy between $\Upsilon_*$ and $\sigma/m$. This degeneracy is expected. Both probes used here (lensing and kinematics) are indifferent to any distinction between baryonic and dark mass, and measure the total mass present. A high stellar mass atop a cored DM density may resemble a low stellar mass in a steep, cuspy DM density profile, yielding this degeneracy. Therefore, our method's ability to distinguish baryonic and dark mass is limited. Better knowledge of $\Upsilon_*$ in central cluster galaxies may therefore help constrain the inner DM slope.

In this work, the lensing and kinematics data vectors provided similar data degrees of freedom (d.o.f.), at 25 and 38 respectively. Pixel-level modeling of strong lensing features, by contrast, typically uses data vectors of hundreds of points or more. \citet{Wang.Canameras.ea2022a}, which also studied a strong lensing cluster, notes this as a potential difficulty in combining strong lensing and kinematics measurements. Combining lensing data with hundreds of data d.o.f., with kinematics likelihoods of typically a few dozen d.o.f., can therefore yield minimal change over a lensing-only model, as the $\chi^2$ of the lensing data will typically dwarf the $\chi^2$ of the much smaller kinematics data vector. \citet{Wang.Canameras.ea2022a} therefore do not combine stellar kinematics with pixel-level image modeling, and only present joint posteriors for a conjugate point likelihood with kinematics, similar to what was done here. Pixel-level SL modeling is beyond the scope of the present work, but we note this as a consideration for future work combining kinematics and strong lensing.

As an additional test, we sampled a variation on our model which explicitly represented the mass of a nearby bright galaxy in Section~\ref{sec:results_external_shear}. While this choice yielded minimal change to our key parameters (chiefly $\Upsilon_*$, $M_{\text{200m}}$, $c_{\text{200m}}$, and $\sigma/m$), implying our results are robust, our likelihood nonetheless strongly preferred the presence of another halo. Future precision models of \MACSJ\, may benefit from explicitly including this halo.

\subsection{\label{sec:comparison_acebron_ertl}Consistency with Other Mass Models}

\begin{table}[]
    \centering
    \renewcommand\arraystretch{1.15}
    \begin{tabular}{c|cc|cc}
        \multirow{2}{3.5em}{Feature} & Lensing & Lensing & Joint & Joint \\
        & Best-Fit & Posterior & Best-Fit & Posterior \\
        \hline
% JOD: this was generated in notebook 2025-08-12 01 RMS stuff
        SN-E & 0.12 & $0.11^{+0.08}_{-0.07}$ & 0.09 & $0.11^{+0.07}_{-0.06}$ \\
SN-R & 0.12 & $0.14^{+0.13}_{-0.07}$ & 0.05 & $0.12^{+0.10}_{-0.06}$ \\
A & 0.20 & $0.21^{+0.16}_{-0.10}$ & 0.24 & $0.21^{+0.14}_{-0.10}$ \\
B & 0.17 & $0.17^{+0.15}_{-0.09}$ & 0.17 & $0.17^{+0.13}_{-0.09}$ \\
C & 0.01 & $0.05^{+0.03}_{-0.02}$ & 0.03 & $0.05^{+0.03}_{-0.02}$ \\
F & 0.04 & $0.06^{+0.04}_{-0.03}$ & 0.06 & $0.05^{+0.04}_{-0.03}$ \\
G & 0.08 & $0.11^{+0.07}_{-0.04}$ & 0.08 & $0.10^{+0.05}_{-0.04}$ \\
H & 0.08 & $0.10^{+0.07}_{-0.04}$ & 0.05 & $0.09^{+0.05}_{-0.03}$ \\
J & 0.06 & $0.09^{+0.07}_{-0.04}$ & 0.03 & $0.08^{+0.05}_{-0.03}$ \\
bulge & 0.10 & $0.20^{+0.10}_{-0.07}$ & 0.08 & $0.20^{+0.10}_{-0.06}$ \\
\hline
all & 0.11 & $0.16^{+0.06}_{-0.04}$ & 0.11 & $0.16^{+0.05}_{-0.03}$ \\
    \end{tabular}
    \caption{\textbf{Image-plane RMS:} The RMS distance between observed and predicted positions of each feature group. The last row gives the combined RMS for all features. All entries are in arcseconds. The `Best-Fit' columns give the result of the maximum likelihood sample.}
    \label{tab:rms-table}
\end{table}

%Early work modeling the strong lensing in \MACSJ\, included \citet{Newman.Belli.ea2018} and \citet{Rodney.Brammer.ea2021}, which presented the discovery of SN Requiem. More recent models include \citet{Acebron.Bergamini.ea2025,Ertl.Suyu.ea2025}, which leverage deep JWST and MUSE data and include SN Encore. Below we compare our results with these previous models. While our results strongly disagree with the model presented in \cite{Rodney.Brammer.ea2021}, they appear broadly consistent with the newest mass models \cite{Acebron.Bergamini.ea2025,Ertl.Suyu.ea2025}.

Shortly after the completion of this work, \citet{Suyu.Acebron.ea2025} and \citet{Pierel.Hayes.ea2025} reported constraints on $H_0$ from both lensed SNe in this system, combining seven independent models using six different codes. Those models included both recent analyses referenced in this work \cite{Ertl.Suyu.ea2025,Acebron.Bergamini.ea2025}. Additionally, \citet{Pierel.Hayes.ea2025} reported a measurement of the time delay between SN Encore images 1a and 1b of $\Delta t_{1b,1a} = -39.8^{+3.9}_{-3.3}$ days, consistent with our model predicted value of $\Delta t_{1a,1b} = -\Delta t_{1b,1a} = 41 \pm 4$ days (see Table~\ref{tab:time-delays}). Note that our prediction was computed with a fixed $H_0 = 67.7$ km/s/Mpc; a higher expansion rate would lower $|\Delta t|$.

Here we compare our results with other published mass models. Common diagnostics used in cluster-scale strong lensing include the critical curves (the contours of infinite magnification), and the root-mean-square scatter between predicted and observed images, commonly called the RMS. The critical curves predicted by our joint model were presented above in Figure~\ref{fig:critical-curves-joint}. While the critical curves can only enable a qualitative comparison, we note that ours appear very consistent with the newest mass models, see Fig. 1 of \cite{Acebron.Bergamini.ea2025} and Fig. 8 of \cite{Ertl.Suyu.ea2025}. Additionally, we present the image-plane RMS of all features used to constrain our model in Table~\ref{tab:rms-table}. We caution against direct comparisons of RMS between our work and other mass models, as we use an entirely different set of position constraints and sample a different likelihood. Nevertheless, both our lensing-only and joint models yield a small, best-fit RMS of 0.11 arcsec.

 Reliable estimates of the lensing magnification are essential to any study of lensed sources, such as \cite{Newman.Gu.ea2025}, and for accurate photometric classification of the observed SNe \cite{Rodney.Brammer.ea2021}.  Our joint model magnifications are in clear agreement with those reported in \citet{Suyu.Acebron.ea2025}, including  \cite{Acebron.Bergamini.ea2025,Ertl.Suyu.ea2025}, with the magnifications of all observed sources within the $1\sigma$ errors of those in \cite{Acebron.Bergamini.ea2025}, and slightly below the absolute magnifications predicted by \cite{Ertl.Suyu.ea2025}. See Fig. 6 of \cite{Acebron.Bergamini.ea2025} and Table 9 of \cite{Ertl.Suyu.ea2025} for comparison. The absolute magnifications $|\mu|$ for predicted images 1d* and 1e* of SN Encore are nearly identical to \cite{Acebron.Bergamini.ea2025} as well\footnote{Curiously, \cite{Acebron.Bergamini.ea2025} lists the demagnified central image 1e* as having opposite sign, with a negative magnification. We consider this to likely be a mistake, demagnified central images always have positive parity.}. Compared to \cite{Ertl.Suyu.ea2025}, we predict $|\mu|$ approximately 20\% lower for the observed image groups (a,b,c). Our predictions for (d,e) are more discrepant, where our predictions are 60\% higher and $\sim$2-5x lower respectively.

 Our results, however, disagree significantly with the earlier model in \citet{Rodney.Brammer.ea2021}, which gave $|\mu|$ of observed images roughly 3-5x lower than ours, and a significantly longer time-delay. \citet{Acebron.Bergamini.ea2025} similarly noted that their predicted magnifications strongly disagree with previously reported lens models in \cite{Rodney.Brammer.ea2021,Newman.Belli.ea2018}. We do not consider this problematic, as we believe our mass model to be higher quality than that in \cite{Rodney.Brammer.ea2021}, which predicted an unreasonably large ellipticity for the overall halo and a BCG more massive than the overall halo. 
Intriguingly, our results in this work also imply a very different time-delay baseline from \cite{Rodney.Brammer.ea2021}, which predicted a roughly 20-year delay between the observed images of SN Requiem and the next predicted image, placing the recurrence in the mid-2030s. Our model yields a much lower delay of roughly $\sim$12 years, in which case SN Requiem would be seen again in the next few years, likely before 2030 (see Table~\ref{tab:time-delays}). Our time-delay prediction is broadly consistent with the meta-analysis in \citet{Suyu.Acebron.ea2025}, which predicts SN Requiem will appear in $\sim$April-December 2026 for $H_0=73$ km s$^{-1}$ Mpc$^{-1}$, or $\sim$March-November 2027 for $H_0=67$ km s$^{-1}$ kpc$^{-1}$.%.Unfortunately, we are unable to compare these predictions to the newer models in \cite{Acebron.Bergamini.ea2025, Ertl.Suyu.ea2025}, which did not report forecasted time delays.

We stress that our group has independently measured the Faber-Jackson relation of the members of \MACSJ\,\cite{Flowers.ODonnell.ea2025}, chosen cluster members to include in the SL mass model, identified multiply imaged features to use as constraints, and uses an entirely different modeling code and parametrized mass model from other work on \MACSJ. While agreement between our predicted magnifications and those in \cite{Acebron.Bergamini.ea2025,Ertl.Suyu.ea2025} does not exclude systematic effects distorting the true magnifications, it is encouraging that these models agree with ours despite substantial differences in approach. \citet{Acebron.Bergamini.ea2025} suggests that the discrepancy between their reported magnifications and earlier work could be partly explained by the inclusion of newly identified sources at different redshift. However, this is unlikely to be true, as our model agrees with theirs despite not using the newly discovered sources. Lens modeling choices still have a large effect on predicted magnifications near the critical curves: our model agrees more closely with \citet{Acebron.Bergamini.ea2025} (which similarly used one cluster halo and an external shear) than it does with \citet{Ertl.Suyu.ea2025} (which used two halos to describe the cluster DM, and no external shear).

\section{\label{sec:conclusion}Conclusions}

\begin{figure}
    \centering
    \includegraphics[width=1\linewidth]{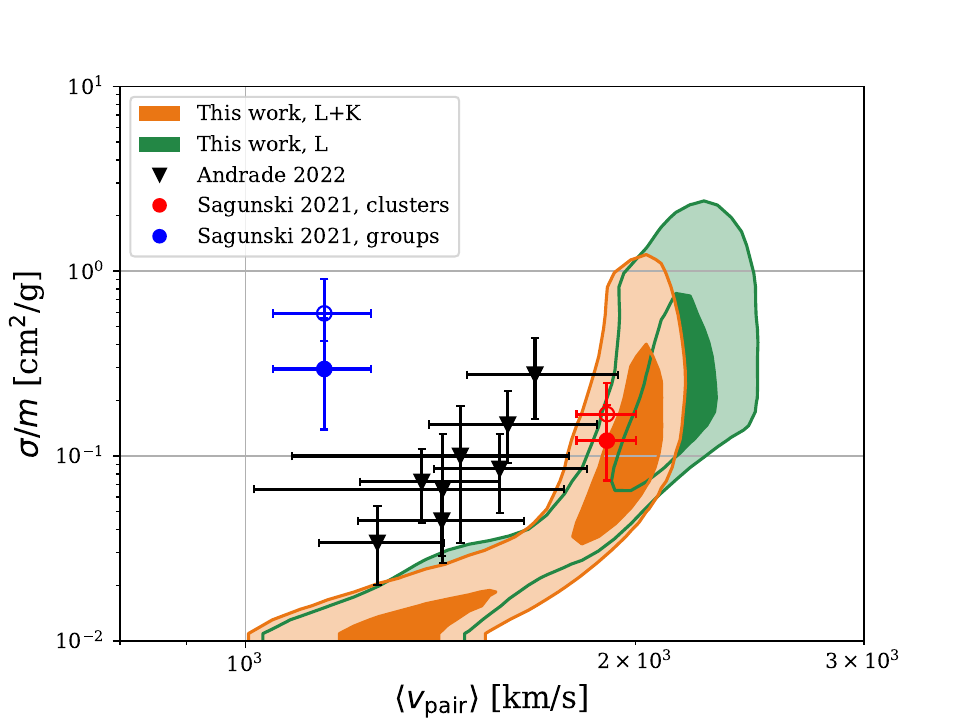}
    \caption{\textbf{SIDM Interaction:} Constraints on the magnitude and velocity scale of SIDM interactions, compared to previous work. Black triangles show the eight clusters studied in \cite{Andrade.Fuson.ea2022}, while the blue and red points show the full constraints from eight groups and seven clusters in \cite{Sagunski.Gad-Nasr.ea2020}. Compare to Figure 3 of \cite{Sagunski.Gad-Nasr.ea2020}.}
    \label{fig:cross-section-v-velocity}
\end{figure}

We have presented a detailed mass model of \MACSJ, combining strong gravitational lensing and stellar kinematics measurements. With the aim of refining previous SIDM constraints on cluster scales, we have improved on the methodology of previous studies \cite{Andrade.Fuson.ea2022,Sagunski.Gad-Nasr.ea2020}. From our combined model, we report a 95\% CL upper limit on the SIDM cross section of $\sigma/m< 0.613$ cm$^2$/g, at an interaction velocity $\langle v_{\text{pair}} \rangle < 2090$ km/s. Though in this work, we studied a single system, we aim to apply this methodology to a sample of systems in the future. Further, these results and methods are of significant interest beyond SIDM. Effectively combining strong lensing and stellar kinematics is widely acknowledged as key to advancing the state-of-the-art of SL modeling. The inclusion of kinematic measurements has the potential to break the well-known mass sheet degeneracy, improving SL-derived cosmological constraints, particularly for time delay cosmography. The \MACSJ\, system is of intense interest, as the first known SL system to host multiple strongly lensed SNe; these results inform future constraints on the Hubble constant from this system.

Our combined kinematics and lensing model places considerably tighter constraints on SIDM interactions than either measurement alone. Our posteriors on the magnitude and velocity scale of self-interaction are shown in in Figure~\ref{fig:cross-section-v-velocity}, with results from \cite{Sagunski.Gad-Nasr.ea2020,Andrade.Fuson.ea2022} overlaid for comparison. For a detailed list of posteriors on measured parameters, along with derived quantities of interest, the reader is referred to Table~\ref{tab:results}. The lensing results alone exhibit a bimodality in $\sigma/m$, with a marked preference for the higher mode near $\sigma/m = 1$ cm$^2$/g. The inclusion of kinematics constraints suppresses this mode.  Notably, our joint results still exhibit a clear bimodality, with one mode at high values of $\Upsilon_*$, $\sigma/m$, and $\langle v_{\text{pair}}\rangle$, and one at lower values of those three parameters. This degeneracy resembles that of the aggregate sample of clusters in \citet{Andrade.Fuson.ea2022}, though offset in $\langle v_{\text{pair}} \rangle-\sigma/m$ space. Improved knowledge of the true stellar mass-to-light ratio of BCGs could therefore break this degeneracy, providing tighter constraints on the inner dark matter halo. %\reviewer{In Fig. 10, the first parameter (the stellar mass-to-light parameter) does not
%have any bimodality for lensing-only and kinematics-only constraints. It is
%counterintuitive to notice that there exists a bimodality when both are
%Combined.
%}

We also note that there is good agreement between our results and those of \cite{Sagunski.Gad-Nasr.ea2020,Andrade.Fuson.ea2022}. While there is some velocity offset, each constraint is assumed to be velocity-independent locally, but all constraints are consistent with the typical $1/v^{4}$ global velocity-dependent behavior of SIDM theory curves in this large velocity limit \cite{Tulin.Yu2018a}.

Extending the methodology in this work to a sample of clusters would greatly improve constraints on SIDM. However, challenges remain. The selection effects inherent in such a sample remain a concern, as strong lensing halos are preferentially concentrated and elongated along the line-of-sight. Fortunately, other probes of the cluster density profile exist with very different systematics, such as X-ray observations of the hot intracluster medium (ICM) and weak lensing measurements. The density profile inferred from X-ray observations of the ICM could be particularly powerful for SIDM studies, as it constrains the cluster halo at similar radii as SL. Weak lensing alone is unlikely to provide meaningful constraints on SIDM \cite{Robertson.Huff.ea2023}, but is highly useful in combination with the other probes here to ensure the inferred cluster mass is accurate on larger scales.

We additionally highlight our forecast for the next predicted images of SNe Requiem and Encore. The study which discovered SN Requiem \cite{Rodney.Brammer.ea2021} predicted a $\sim$20 year time delay, with the next images appearing in the mid to late 2030s. Our results differ significantly, implying a baseline of roughly 12 years, with SNe Requiem and Encore likely reappearing in 2027-2028 and 2033-2034, respectively. These predictions use a fixed cosmology with H$_0 = 67.7$ km s$^{-1}$ Mpc$^{-1}$; a larger expansion rate would shorten these time delays. Observations of these images would enable a time-delay measurement of the Hubble constant of unprecedented precision from a single system. Monitoring for the reappearance of SN Requiem could therefore provide enormous benefit to cosmology sooner than previously expected.

% I say "the \MACSJ cluster" here to distinguish from Andrew Newman's recent paper, doing spatially resolved kinematics on the source.
This work is the first to leverage spatially resolved stellar kinematics to model the \MACSJ\, cluster. While the kinematics map measured in this work was computed from MUSE observations, there exist newer, deeper MUSE observations on this system. Future work leveraging these data has enormous potential, especially given recent progress in rigorously characterizing the use of stellar kinematics for time-delay lenses \cite{Shajib.Mozumdar.ea2023,Knabel.Mozumdar.ea2025,Shajib.Treu.ea2025}. 
Our model contributes to the growing body of published mass models on \MACSJ\,\cite{Rodney.Brammer.ea2021,Newman.Belli.ea2018,Acebron.Bergamini.ea2025,Ertl.Suyu.ea2025}, a remarkable lens system enabling an enormous variety of science. %Further comparison of our mass model with previous studies of \MACSJ\, can be found in Appendix~\ref{appendix:sne-fits}.

%{\color{red} Thoughts and suggestions for future models of \MACSJ. Incorporating X-ray data; Modeling with that other nearby mass; Pixel-level constraints; Using weak-lensing maps of the field.}

\begin{acknowledgments}
We gratefully acknowledge helpful correspondence and feedback from Andrew Robertson, Anowar Shajib, Michele Cappellari,  Ana Acebron, and Sherry Suyu.

% JOD: citation per instruction here https://archive.eso.org/cms/eso-data-access-policy.html#acknowledgement
This work utilized data from the ESO Science Archive Facility with DOI: https://doi.eso.org/10.18727/archive/41.

This work made extensive use of standard, open-source Python packages. In addition to those noted throughout the text, these include \texttt{astropy} \cite{astropy-01,astropy-02,astropy-03}, \texttt{numpy} \cite{numpy}, \texttt{scipy} \cite{scipy}, and \texttt{matplotlib} \cite{matplotlib}. Additionally, we acknowledge use of the lux supercomputer at UC Santa Cruz, funded by NSF MRI grant AST 1828315.
\end{acknowledgments}

\appendix

\section{\label{appendix:jeans}Isothermal Jeans modeling}

To describe the effect of SIDM on the cluster density profile, we utilize the isothermal Jeans formalism briefly described above in Section~\ref{sec:methods_sidm_parameters}. This results in a piecewise DM density profile:

\begin{equation}
    \rho_{\text{SIDM}}(r) = \begin{cases}
        \rho_{\text{iso}}(r) & r < r_1 \\
        \rho_{\text{NFW}}(r) & r > r_1
    \end{cases}
\end{equation}

Where $\rho_{\text{NFW}}$ denotes the usual Navarro-Frenk-White density profile, and $\rho_{\text{iso}}$ describes the thermalized region, obeying the isothermal Jeans equation with an isothermal velocity of $\sigma_0$.
We make the same assumption as in \citep{Robertson.Massey.ea2021} and \citep{Sagunski.Gad-Nasr.ea2020} that the expected pairwise particle velocity is $\langle v_{\text{pair}}\rangle = \frac{4\,\sigma_0}{\sqrt{\pi}}$. The SIDM cross sections in this work are reported as $\sigma/m = \langle\frac{\sigma v}{m}\rangle \langle v_{\text{pair}}\rangle^{-1}$ (c.f. \citet{Sagunski.Gad-Nasr.ea2020} eq. 6). To ensure consistency with the outer NFW halo, boundary conditions are enforced at $r_1$: the enclosed mass $M(<r_1)$ and density $\rho(r)$ are identical to an NFW halo.

We use a similar change of variables as in Section 2.2 of \citet{Robertson.Massey.ea2021}, modified to account for the contribution of a baryonic density profile $\rho_b(r)$. In this work, we use the stellar density profile of the BCG as $\rho_b(r)$, inferred from the the MGE approximation of the BCG and $\Upsilon_*$. From the Jeans equation for a thermalized DM profile $\rho_{\text{iso}}$,

\begin{equation}
    \frac{d}{dr} \Big(r^2\frac{d\ln\rho_{\text{iso}}}{dr}\Big) = - \frac{4\pi G r^2 (\rho_{\text{iso}} + \rho_b)}{\sigma_0^2}
\end{equation}

we introduce the change of variables $x = r/r_1$, $y = \ln(\rho_{\text{iso}}/\rho_0)$, $N_0 = \rho_0 / \rho_1$, yielding:

\begin{align}
    \frac{d^2y}{dx^2} &= -\frac{2}{x}\frac{dy}{dx} - C\Big(e^y + \frac{\rho_b(xr_1)}{\rho_0}\Big) \\
    C &= \frac{4\pi G \rho_1 r_1^2 N_0}{\sigma_0^2}
\end{align}

To numerically solve the equation above for $\rho_{\text{iso}}(r)$, we use the $\texttt{scipy}$ function $\texttt{solve\_bvp}$. This solver simultaneously fits the above ODE and the free parameters $N_0$ and $\sigma_0$, subject to the following boundary conditions:

\begin{align*}
    y|_{x=0}&=0 & M|_{x=0} &= 0 \\
    \frac{dy}{dx}|_{x=0}&= 0 & y|_{x=1} &= -\ln(N_0) \\
    M|_{x=1}&= M_\text{NFW}(<r_1) \\
\end{align*}

As may be apparent from the discussion above, the SIDM cross section $\sigma/m$ is not directly sampled. Rather, the SIDM parameters are computed from $\Upsilon_*$, $M_{200m}$, $c_{200m}$, and the radius $r_1$. (This sampling via $r_1$ is termed the `outside-in' method in \citet{Robertson.Massey.ea2021}, as opposed to the `inside-out' method computing directly from $\sigma/m$.) This poses a problem: since $\sigma/m$ is a derived parameter, we can only place a prior on $r_1$, rather than $\sigma/m$. We desire a flat prior on $\ln(\sigma/m)$, which would imply the following prior on $\ln r_1$:

\begin{align}
    P(\ln r_1) = \frac{d(\sigma/m)}{dr_1} \frac{r_1}{\sigma/m}
\end{align}

Though we sample from a flat prior $P(\ln r_1) = \text{const}$, we enforce our desired prior by adding a weighting $\ln\Big(\frac{d(\sigma/m)}{dr_1} \frac{r_1}{\sigma/m}\Big)$ to the log likelihood.

\section{\label{appendix:triaxiality}Triaxiality and Concentration Bias}

While it is well known that strong lensing selected halos are biased to high concentrations, they additionally exhibit a severe triaxial bias: halos elongated along the line-of-sight are more likely to exhibit strong lensing. In addition, this triaxial bias is likely to artificially inflate the apparent concentration of lens halos, potentially as much as $+60\%$ (\citet{Andrade.Fuson.ea2022}, Appendix C).

We allow for an elliptical halo in three dimensions, rather than limiting ellipticity to the plane of the sky. We modify both our strong lensing and stellar kinematics models accordingly. The majority of cluster halos in the universe are expected to be roughly prolate \cite{Bonamigo.Despali.ea2015}; as such, we allow for a prolate, rather than oblate or fully triaxial ellipticity. (In addition, the axisymmetric solutions of \texttt{JAM}, used for kinematics modeling, only allow for oblate or prolate halos, not fully triaxial.)

%{\color{red}JOD: This paragraph is an overly-polemical rough draft. Refine and make nicer.}

The introduction of a 3D ellipticity to matter halos is rarely treated with the subtlety it deserves. \citet{Andrade.Fuson.ea2022} marginalizes over LOS ellipticity by inflating the lensing convergence $\kappa_e (\theta)= \kappa(\theta)/s$, for an axis ratio $s<1$. While this correctly inflates the observed lensing, the mass and concentration of the halo are no longer meaningful: the apparent lensing now corresponds to a halo of mass $M/s$. Likewise, a common approach takes a spherical density profile $\rho_{\text{sph}}(r)$, and introduces an elliptical radius $r_e^2 = \frac{x^2}{q^2} + \frac{y^2}{q^2} + z^2$ such that $\rho_e(\vec{x}) = \rho_{\text{sph}}(r_e)$.\footnote{For $q<1$, this corresponds to a prolate (`pickle'-shaped) halo, and for $q>1$ this represents an oblate (`pancake'-shaped) halo.} % {\color{red} TODO who actually uses this? cite}
However, this similarly fails to preserve important properties of the halo: the total halo mass is now $M_e = q^2 M_{\text{sph}}$.

In order to preserve the spherical overdensity (SO) halo definition for elliptical halos, and therefore the meaning of the mass and concentration $(M_{\Delta}, c_{\Delta})$, we introduce ellipticity as follows. A SO halo of volume $V_{\Delta} = \frac{4\pi}{3} R^3_{\Delta}$ has a mass of $M_{\Delta} = \Delta V_{\Delta} \rho_{b}(z)$, where $\rho_b(z)$ is a background density, commonly taken to be either the mean matter density $\rho_m(z)$ or the critical density of the universe $\rho_c(z)$, and $\Delta$ is the overdensity of the halo within $R_{\Delta}$ compared to this reference. We wish to preserve $V_{\Delta,e} = V_{\Delta,\text{sph}}$ and $M_{\Delta,e} = M_{\Delta,\text{sph}}$. For a prolate halo with axis ratio $q_{3d}<1$, this requires:

\begin{align}
    r_e^2 &= \frac{x^2}{q^{2/3}} + \frac{y^2}{q^{2/3}} + q^{4/3}z^2 \\
    \rho_e(\vec{x}) &= \rho_{\text{sph}}(r_e) 
\end{align}

%\grantcomment{why are the q values for Eq. 1 now having different powers? I don't see how this follows exactly.} \jackcomment{It's just scaling $r_e$ to preserve the mass and volume of the halo. The more usual $(x/q)^2 + (y/q)^2 + z^2$ shrinks the halo volume and mass by $q^2$.} \grantcomment{I get the concept, I'm just a bit confused how these extra factors of q are exactly determined. I don't see how to directly calculate that this is the right factor. There's a lot of subscripts that also make it a little unclear. }
Where the major axis is aligned with the $z$-axis. We introduce an angle $i$ between $z$ and the LOS, such that $z^\prime$ becomes the true LOS and the observed coordinates on the sky are $(x^\prime, y^\prime)$. Strong lensing, however, is dependent on the projected mass density $\Sigma(x^\prime,y^\prime) = \int \rho(x^\prime,y^\prime,z^\prime)dz^\prime$. We will use lowercase $r$ to represent spherical 3D radii, and distinguish projected 2D radii with a capital $R$. The projected mass density of our prolate halo is therefore:

\begin{align}
    \label{eq:projected_Sigma}
    \Sigma_e(R_e) &= \frac{q_{3d}^{1/3}}{\sqrt{\sin^2i + q_{3d}^2\cos^2i}} \Sigma_{\text{sph}}(R_e) \\
    \label{eq:projected_re}
    R_e^2 &= \frac{x^{\prime2}}{q_{3d}^{2/3}} + \frac{q_{3d}^{4/3} y^{\prime2}}{\sin^2i + q_{3d}^2 \cos^2i} \\
    \label{eq:projected_ellipticity}
    q_{2d} &= \frac{q_{3d}}{\sqrt{\sin^2i + q_{3d}^2\cos^2i}}
\end{align}

Where $\Sigma_{\text{sph}}$ is the projected density for a spherical halo, and $q_{2d}$ is the apparent projected ellipticity.\footnote{Compare Eq.~\ref{eq:projected_ellipticity} to \citet{Cappellari2020} Eq. 36.} For any basis function for a spherically symmetric, 3D density $\rho(r; s)$ with a scale radius $s$, and corresponding projected density $\Sigma_{\text{sph}}(R; s)$, 3D ellipticity can thus be trivially introduced by Eqs. \ref{eq:projected_Sigma}-\ref{eq:projected_ellipticity}. We take advantage of this property to approximate our 3D, elliptical DM halo as a sum of Cored Steep Ellipsoids (CSEs) or Gaussians, in order to model the observed lensing and dynamics, respectively. See Sections ~\ref{sec:methods_strong_lensing} and \ref{sec:methods_stellar_kinematics}.

\section{\label{appendix:kinematics_systematics}Robustness of the Measured Stellar Kinematics}

\begin{figure}
    \centering
    \includegraphics[width=1.0\linewidth]{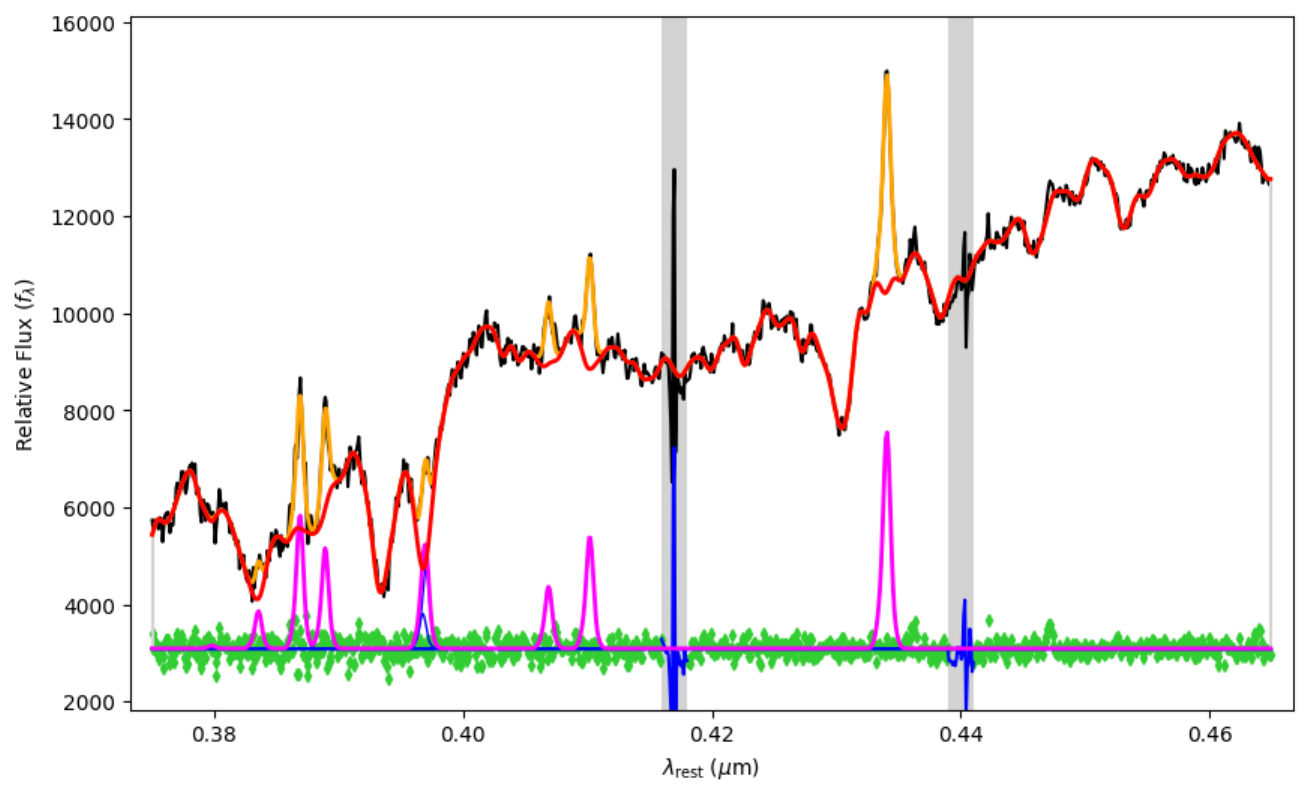}
    \includegraphics[width=1.0\linewidth]{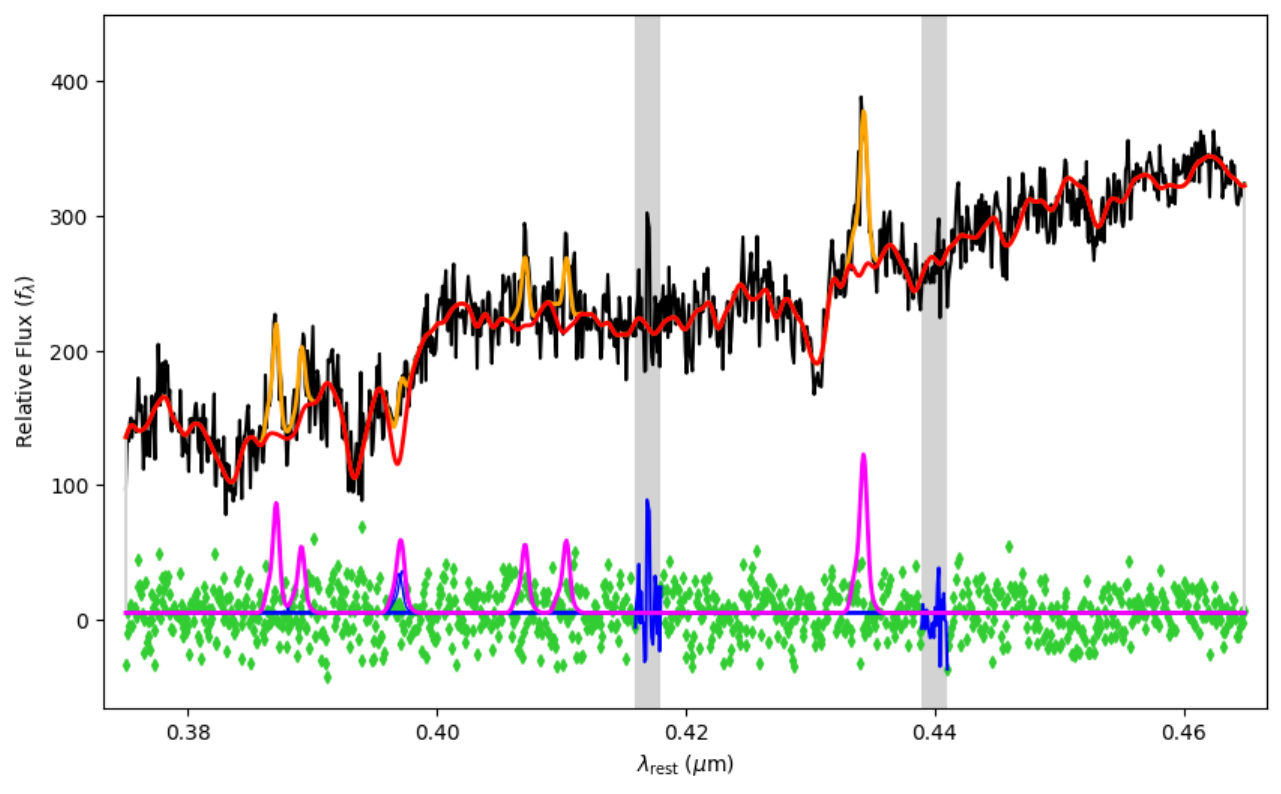}
    \caption{Example \texttt{ppxf} fit results. Black represents data, red represents stellar model fit, orange represents stellar + gas model fit. \textbf{Top:} An example `total' fit to all Voronoi bins combined, used to determine the optimal stellar template. Fit yielded a reduced $\chi^2$ of 1.62. \textbf{Bottom:} An example fit to an individual Voronoi bin. This fit yielded a reduced $\chi^2$ of 1.01.}
    \label{fig:ppxf_fits}
\end{figure}

\begin{figure}
    \centering
    \includegraphics[width=1.0\linewidth]{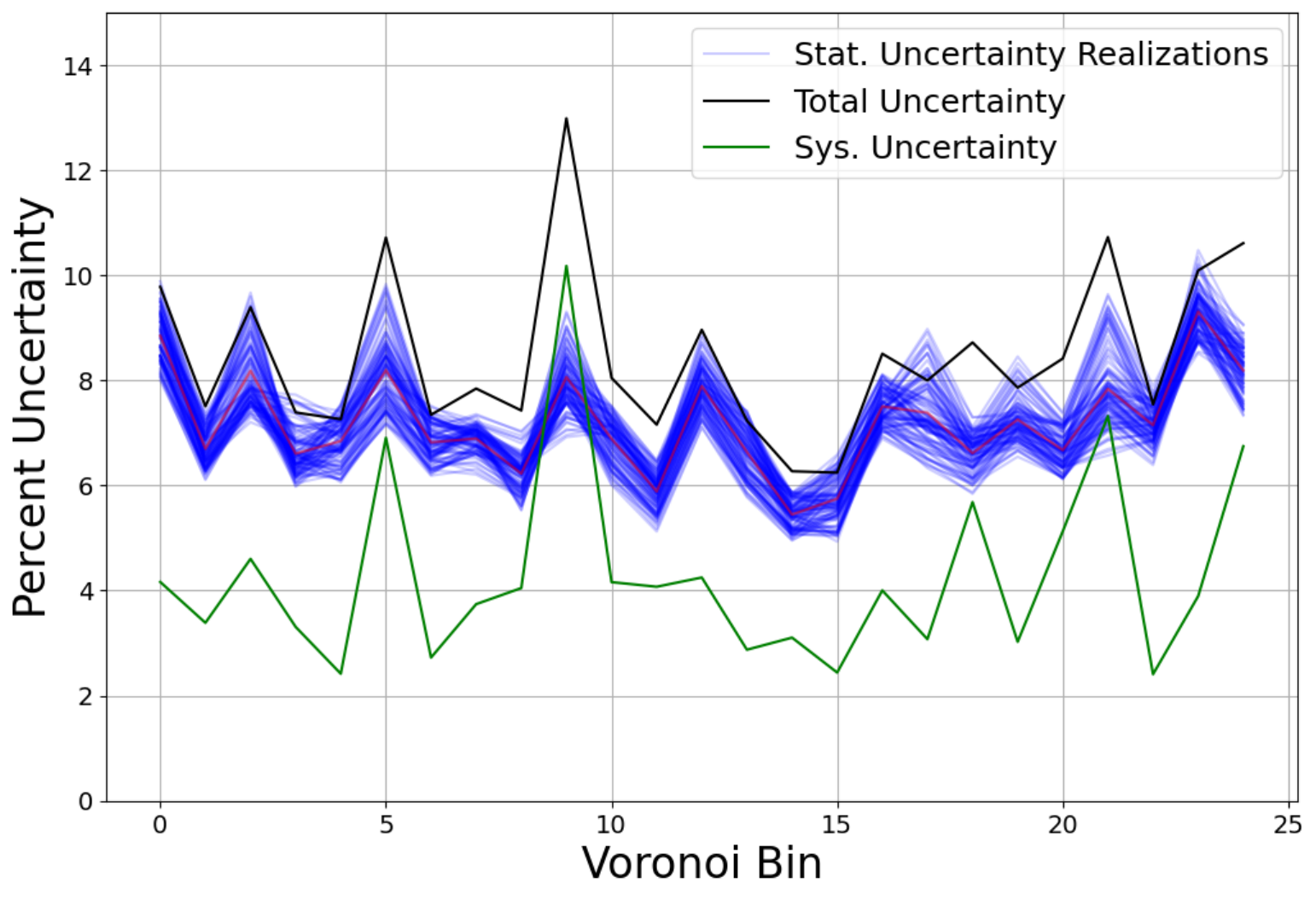}
    \caption{Relative uncertain of the velocity dispersions measured in this work, for each voronoi bin. The total (systematic + statistical) uncertainty is shown in black; the statistical uncertainties of all 108 realizations described in Section~\ref{sec:stellar_kinematics_cov} are shown in blue; and the systematic uncertainty is shown in green.}
    \label{fig:relative-uncertainties}
\end{figure}

Measurements of stellar kinematics can be highly sensitive to analysis choices, and we have drawn on recent work, particularly \cite{Knabel.Mozumdar.ea2025,Shajib.Mozumdar.ea2023}, to ensure our measurements are as robust as possible. To illustrate the quality of our spectral fits, two sample \texttt{ppxf} fits in this work are shown in Figure~\ref{fig:ppxf_fits}. These demonstrate both the overall fit to the entire BCG used to determine an optimal spectral template, and a characteristic fit to an individual Voronoi bin.

The relative uncertainty of the stellar velocity dispersion measured in each Voronoi bin is shown in Figure~\ref{fig:relative-uncertainties}. Total uncertainty is of order $\sim10\%$ in each bin, and systematic uncertainty is subdominant to statistical uncertainty in all bins save for bin 9, the northernmost bin. \citet{Knabel.Mozumdar.ea2025} has rigorously explored systematic uncertainty in similar measurements with the goal of sub-percent precision $\sigma_v$ measurements, and demonstrated that the choice of stellar template library is often the dominant source of systematic uncertainty. In this work we exclusively used the XSL template library, limiting our ability to characterize this uncertainty. However, the same paper demonstrated such an effect on measurements using MUSE spectra was typically of order $\approx 3-4\%$, well below the statistical uncertainty in this work, and of similar scale to the systematic uncertainty we observe. However, our statistical uncertainty is large compared to \cite{Shajib.Mozumdar.ea2023,Knabel.Mozumdar.ea2025} in part because we use a less stringent SNR target inspired by \cite{Newman.Treu.ea2013a}. A higher target spectral SNR threshold would yield fewer Voronoi bins, with lower statistical uncertainties. We conducted several informal tests varying the spatial mask used to Voronoi bin the BCG, and increasing the spectral SNR target: in all cases, the kinematics maps produced were qualitatively consistent with the fiducial map used in this work. %{\color{red} TODO explain that our snr cut differs from Anowar et al.'s}

%{\color{red} Compare chisq maps?}

%\section{\label{appendix:anisotropy}Velocity Anisotropy}

%{\color{red}TODO}

%\section{\label{appendix:sne-fits}Magnifications of SNe Requiem and Encore}

%\section{\label{appendix:full_posteriors}Full Posteriors}

%{\color{red} Posterior corner plots will be massive, we'll only show the subset of interesting parameters above. Here we'll show the full thing.}

%\nocite{*}

\bibliography{apssamp}% Produces the bibliography via BibTeX.

\end{document}

%% file: position_constraints_table.tex
\renewcommand\arraystretch{1.1}
\begin{tabular}{c >{\centering}p{0.12\linewidth} >{\centering}p{0.12\linewidth} >{\centering}p{0.1\linewidth}cc}
\toprule
    Feature & \multirow{2}{1.5em}{RA} & \multirow{2}{5em}{Declination} & Semi-Major & Semi-Minor & Position \\
     Label & & & Axis & Axis & Angle \\ \midrule[0.1em]
        %H.1 & & & & & \\
        %H.2 & & & & & \\
        % \midrule
        % &  & & & & \\
bulge.1 & 01:38:03.93 & -21:55:49.23 & 0.91" & 0.13" & 75.69\degree \\
bulge.2 & 01:38:03.17 & -21:55:47.69 & 0.58"& 0.11" & -63.68\degree \\
bulge.3 & 01:38:02.38 & -21:55:33.68 & 0.47" & 0.11"& 4.55\degree \\
bulge.4 & 01:38:04.23 & -21:55:24.07 & 0.43" & 0.17" & 46.12\degree \\
bulge.5 & 01:38:03.80 & -21:55:31.09 & 0.12" & 0.07" & 41.44\degree \\
 \hline
SN-E.1 & 01:38:03.75 & -21:55:49.66 & 0.1" & --- & --- \\
SN-E.2 & 01:38:03.17 & -21:55:47.85 & 0.1" & --- & --- \\
\hline
SN-R.1 & 01:38:03.63 & -21:55:50.40 & 0.15" & --- & --- \\
SN-R.2 & 01:38:02.96 & -21:55:47.23 & 0.15" & --- & --- \\
SN-R.3 & 01:38:02.42 & -21:55:38.43 & 0.15" & --- & --- \\
\hline
A.1 & 01:38:03.48 & -21:55:50.29 & 0.1" & --- & --- \\
A.2 & 01:38:03.13 & -21:55:48.74 & 0.1" & --- & --- \\
A.3 & 01:38:02.40 & -21:55:38.62 & 0.1" & --- & --- \\
\hline
    B.1 & 01:38:03.65 & -21:55:50.16 & 0.1" & --- & --- \\
B.2 & 01:38:03.08 & -21:55:47.92 & 0.1" & --- & --- \\
B.3 & 01:38:02.39 & -21:55:37.15 & 0.1" & --- & --- \\
\hline
C.1 & 01:38:03.89 & -21:55:49.67 & 0.1" & --- & --- \\
C.2 & 01:38:03.06 & -21:55:47.31 & 0.1" & --- & --- \\
\hline
F.1 & 01:38:04.04 & -21:55:48.83 & 0.1" & --- & --- \\
F.2 & 01:38:03.10 & -21:55:47.12 & 0.1" & --- & --- \\
\hline
G.1 & 01:38:04.16 & -21:55:48.20 & 0.1" & --- & --- \\
G.2 & 01:38:03.07 & -21:55:46.46 & 0.1" & --- & --- \\
G.3 & 01:38:02.40 & -21:55:32.20 & 0.1" & --- & --- \\
\hline
H.1 & 01:38:04.17 & -21:55:47.86 & 0.1" & --- & --- \\
H.2 & 01:38:03.14 & -21:55:46.76 & 0.1" & --- & --- \\
H.3 & 01:38:02.40 & -21:55:30.69 & 0.1" & --- & --- \\
\hline
J.1 & 01:38:04.10 & -21:55:48.16 & 0.1" & --- & --- \\
J.2 & 01:38:03.20 & -21:55:47.23 & 0.1" & --- & --- \\
J.3 & 01:38:02.39 & -21:55:30.96 & 0.1" & --- & --- \\
         \bottomrule
    \end{tabular}

%% file: results_table.tex
\centering
    \renewcommand{\arraystretch}{1.2}
    % JOD: This magic incantation makes 6 columns, each centered and aligned vertically to the middle, with custom widths
    % There is a dummy 0-width 7th column because mysterious things break without it :shrug:
    \begin{tabularx}{1\linewidth}{>{\centering}m{0.12\linewidth} m{0.4\linewidth} >{\centering}m{0.1\linewidth} >{\centering}m{0.12\linewidth} >{\centering}m{0.12\linewidth} >{\centering}m{0.12\linewidth} m{0.0\linewidth}}
    \toprule
    \multirow{2}{5em}{Parameter} & \centering \multirow{2}{5em}{Description} & \multirow{2}{2.5em}{Prior} & Lensing & Kinematics & Joint &\\
    & & & Result & Result & Result &\\
    \midrule
    \multicolumn{2}{l}{Shared Parameters} & & & & &\\%[-3ex]
    \hline
    $\log_{10}\Upsilon_*$ & BCG stellar mass-to-light ratio & $\mathcal{N}(0.31, 0.3)$ & $0.50^{+0.12}_{-0.21}$ & $0.40^{+0.21}_{-0.29}$ & $0.47^{+0.19}_{-0.35}$ & \\
    $\log_{10}(\frac{M_{200m}}{M_\odot})$ & Overall halo NFW mass & $\mathcal{U}(13.5, 15.5)$ & $14.90^{+0.20}_{-0.17}$ & $14.52^{+0.61}_{-0.59}$ & $14.81^{+0.14}_{-0.12}$ & \\
    $c_{200m}$ & Overall halo NFW concentration & $\mathcal{U}(5, 15)$& $10.14^{+2.11}_{-3.03}$  & $10.73^{+2.93}_{-3.41}$ & $9.92^{+2.19}_{-1.88}$ & \\
    $\log_{10}(r_1/\text{kpc})$ & Isothermal Jeans radius, defines SIDM behavior & $\mathcal{U}(0.0, 3)$& $2.28^{+0.28}_{-0.92}$ & $1.69^{+0.60}_{-0.95}$ & $1.58^{+0.69}_{-0.99}$ & \\
    $q$ & 3D halo prolate axis ratio & $\mathcal{U}(0.3, 1)$& $0.64^{+0.05}_{-0.10}$ & $0.67^{+0.22}_{-0.24}$ & $0.61^{+0.05}_{-0.13}$ & \\
    $\cos(i)$ & Halo inclination relative to LOS & $\mathcal{U}(0, 1)$ & $0.44 \pm 0.29$ & $0.73^{+0.18}_{-0.40}$ & $0.51^{+0.28}_{-0.32}$ & \\
    $\phi$ & Position angle of projected ellipticity in the plane of the sky. East-of-North & $\mathcal{U}(-48\degree, 132\degree)$ & $44.3^{+1.8}_{-1.7}$ & $42.0^{+43.1}_{-43.3}$ & $43.6^{+1.6}_{-1.5}$ & \\
    \hline
    \multicolumn{2}{l}{Lensing Parameters} & & & & & \\%[-3ex]
    \hline
    $x_{\text{halo}}$ & Horizontal offset between halo and BCG center, arcsec & $\mathcal{N}(0.0, 0.3)$ & $-0.10^{+0.22}_{-0.26}$ & --- & $0.02^{+0.16}_{-0.18}$ & \\
    $y_{\text{halo}}$ & Vertical offset between halo and BCG center, arcsec& $\mathcal{N}(0.0, 0.3)$ & $-0.33^{+0.23}_{-0.26}$ & --- & $-0.25^{+0.19}_{-0.21}$ & \\
    $\gamma_{1}$ & External shear component 1 & $\mathcal{N}(0.0, 0.05)$ & $0.02 \pm 0.01$ & --- & $0.02 \pm 0.01$ & \\
    $\gamma_{2}$ & External shear component 2 & $\mathcal{N}(0.0, 0.05)$ & $-0.08^{+0.03}_{-0.02}$ & --- & $-0.06 \pm 0.02$ & \\
    $\alpha$ & Slope of the Faber-Jackson relation & $\mathcal{N}(0.26, 0.06)$ & $0.26 \pm 0.06$ & --- & $0.26 \pm 0.06$ & \\
    $\sigmalt^{\text{piv}}$ & dPIE velocity dispersion scale of Faber-Jackson relation, km/s & $\mathcal{N}(199.1, 22.3)$ & $184.8^{+19.3}_{-20.4}$ & --- & $188.8^{+19.4}_{-19.5}$ & \\
    $\sigmalt^{\text{J1}}$ & dPIE velocity dispersion of J1, km/s & $\mathcal{N}(100, 50)$ & $66.5^{+31.0}_{-33.0}$ & --- & $65.1^{+35.7}_{-33.7}$ & \\
    $\sigmalt^{\text{J3}}$ & dPIE velocity dispersion of J3, km/s & $\mathcal{N}(100, 50)$& $51.0^{+24.9}_{-27.0}$ & --- & $49.6^{+25.3}_{-26.8}$ & \\
    $r_s^{\text{J4}}$ & dPIE velocity dispersion of J4, km/s & $\mathcal{N}(100, 50)$& $19.6 \pm 5.2$ & --- & $19.5^{+5.1}_{-5.2}$ & \\
    $\sigmalt^{\text{J4}}$ & dPIE scale radius of J4, kpc & $\mathcal{N}(20, 5)$& $48.3^{+14.7}_{-16.8}$ & --- & $56.6^{+12.0}_{-14.6}$ & \\
    \hline 
    \multicolumn{2}{l}{Kinematics Parameters} & & & & & \\%[-3ex]
    \hline
    $\betasph$& Velocity anisotropy & $\mathcal{U}(-0.4, 0.4)$& --- & $-0.09^{+0.21}_{-0.19}$ & $-0.15^{+0.15}_{-0.14}$ & \\
    $\sigma_{\text{PSF}}$ & Gaussian PSF size, arcsec & $\mathcal{U}(-0.35, 0.5)$& --- & $0.43 \pm 0.05$ & $0.43 \pm 0.05$ & \\
    \hline 
    \multicolumn{2}{l}{Derived Parameters} & & & & & \\%[-3ex]
    \hline
    $\sigma/m$ & SIDM Cross Section, cm$^2$/g & --- & $0.16^{+0.50}_{-0.15}$ & $0.04^{+0.35}_{-0.03}$ & $0.02^{+0.18}_{-0.02}$ & \\
    $\log_{10}(\sigma/m)$ & Logarithm of $\sigma/m$ in cm$^2$/g & --- & $-0.8^{+0.6}_{-1.1}$ & $-1.4^{+1.0}_{-0.8}$ & $-1.7^{+1.0}_{-0.8}$ & \\
    $\sigma_0$ & Thermal velocity of isothermal region, km/s & --- & $919^{+88}_{-333}$ & $565^{+375}_{-250}$ & $691^{+200}_{-406}$ & \\
    $V_{\text{max}}$ & Maximum circular velocity of cluster DM halo & --- & $1518^{+161}_{-137}$ & $1144^{+640}_{-410}$ & $1422^{+95}_{-82}$ & \\
    $q_{\text{2D}}$ & Projected apparent axis ratio of cluster DM & --- & $0.68 \pm 0.03$ & $0.82^{+0.14}_{-0.28}$ & $0.67 \pm 0.02$ & \\
    $\log_{10}M^{\text{enc}}_{20}$ & Enclosed mass within 20kpc & --- & $12.22^{+0.14}_{-0.16}$ & $12.24^{+0.13}_{-0.19}$ & $12.34^{+0.04}_{-0.05}$ & \\
    $\log_{10}M^{\text{proj}}_{100}$ & Projected mass within 100kpc & --- & $13.79 \pm 0.03$ & $13.61^{+0.29}_{-0.33}$ & $13.78^{+0.02}_{-0.04}$ & \\
    \bottomrule & & & & & & 
    \end{tabularx}